\documentclass[12pt]{article}
\usepackage{amsfonts}
\usepackage{amsmath}
\usepackage{amssymb}
\usepackage{amsthm}
\usepackage{accents}
\usepackage{youngtab}
\usepackage{braket}
\usepackage{graphicx}
\usepackage[svgnames]{xcolor}
\usepackage{ytableau}
\usepackage{here}
\usepackage{comment}
\usepackage{tikz}
 \usetikzlibrary{patterns}
\usepackage{feynmf}
\usepackage{hyperref}
\usepackage{simplewick}
\usepackage{mathrsfs}
\usepackage{latexsym}

\hypersetup{
linktocpage=true
}

\newcommand{\ba}{\begin{eqnarray}}
\newcommand{\ea}{\end{eqnarray}}
\newcommand{\bq}{\begin{equation}}
\newcommand{\eq}{\end{equation}}
\newcommand{\Z}{{\mathbb Z}}
\newcommand{\SU}{{\rm SU}}
\newcommand{\U}{{\rm U}}

\makeatletter
 
  \@addtoreset{equation}{section}
 \makeatother

\begin{document}

\begin{titlepage}

\begin{flushright}
\ 
\end{flushright}

\vskip 12mm

\begin{center}
{\Large{\bf Quantum deformation of Feigin-Semikhatov's W-algebras and 5d AGT correspondence  with a simple surface operator}\par}
\end{center}

\vskip 2cm

\begin{center}
{\Large Koichi Harada}
\vskip 2cm
{\it Department of Physics, The University of Tokyo}\\
{\it 7-3-1 Hongo, Bunkyo-ku, Tokyo 113-0033, Japan}
\end{center}
\vfill
\begin{abstract}
The quantum toroidal algebra of $\mathfrak{gl}_1$ provides many deformed W-algebras associated with (super) Lie algebras of type A. The recent work by Gaiotto and Rap\v{c}\'{a}k suggests that a wider class of deformed W-algebras including non-principal cases are obtained by gluing the quantum toroidal algebras of $\mathfrak{gl}_1$. These algebras are expected to be related with 5d AGT correspondence.  In this paper, we discuss  quantum deformation of the W-algebras   obtained from $\widehat{\mathfrak{su}}(N)$ by the quantum Drinfeld-Sokolov reduction with $\mathfrak{su}(2)$ embedding $[N-1,1]$. They were studied by Feigin and Semikhatov and we refer to them as Feigin-Semikhatov's W-algebras. We construct free field realization and find several quadratic relations. We also compare the norm of the Whittaker states with the instanton partition function under the presence of a simple surface operator in the $N=3$ case. 

\end{abstract}
\vfill
\end{titlepage}

\tableofcontents

\section{Introduction}
\label{sec:intro}
The study of deformed chiral algebras  has begun in 1990s. It was pioneered in \cite{Shiraishi:1995rp},  where quantum deformation of Virasoro algebra was constructed. It has been generalized to the principal W-algebras for type A in \cite{Awata:1996dx} and for simple Lie algebras in \cite{FRqW}. 

 After the discovery of the AGT correspondence \cite{Alday2010,Gaiotto:2009ma,Wyllard2009}, a new way to deal with chiral algebras  was developed \cite{Maulik:2012wi,schiffmann2013cherednik}. In this correspondence, the action of the chiral algebra on the instanton moduli space is described well by $W_{1+\infty}$ or the affine Yangian. 
When the  correspondence is extended to 5d supersymmetric gauge theories \cite{awata2010five}, the dual algebras lift to  deformed W-algebras and they can be understood as truncations of the quantum toroidal algebras. In particular, the quantum toroidal algebra of $\mathfrak{gl}_1$ (also called Ding-Iohara-Miki algebra \cite{Ding:1996mq,Miki2007}) provides the unified way to deal with all of the  deformed W-algebras of type A. It  was also found that it contained the deformed W-algebras q-$W(\widehat{\mathfrak{gl}}_{n|m})$ associated with $\mathfrak{gl}_{n|m}$ \cite{bershtein2018plane}. 
Further, it has been recently clarified that the quantum toroidal algebra of $\mathfrak{gl}_1$ reproduces the deformed W-algebras for the other simple Lie algebras \cite{Feigin:2020glm}.

The CFT limit of q-$W(\widehat{\mathfrak{gl}}_{n|m})$ has been studied also from the different viewpoints \cite{Gaiotto:2017euk,Litvinov:2016mgi,Rapcak:2018nsl}. 
In \cite{Gaiotto:2017euk}, Gaiotto and Rap\v{c}\'{a}k constructed  these algebras in the context of  supersymmetric gauge theoreis. They considered the system consisting of  a 5-brane junction and several D3 branes. The chiral algebra referred to as Y-algebra appears at the 2d corner in this configuration and its explicit form can be described by the quantum Drinfeld-Sokolov reduction and the coset construction. Although the definition is quite different from $W(\widehat{\mathfrak{gl}}_{n|m})$, the character analysis suggests that they are the same algebras. 
 One of the interesting features in Gaiotto-Rap\v{c}\'{a}k's construction is that one can extend the algebras by considering more general brane-webs \cite{Prochazka:2017qum,Prochazka:2018tlo,Creutzig:2017uxh}. 
 There are two ways to define these algebras. In  the same manner as the case of Y-algebra, one can define them by  the quantum Drinfeld-Sokolov reduction and the coset construction. On the other hand, one can also see the system as the combination of the Y-algebras and the operators connecting two junctions. The relation between these two descriptions are known as  so-called module extension. 

All of the non-principal W-algebras of type A can be realized in the above framework. 
In the AGT correspondence, these algebras appear when we consider the supersymmetric gauge theories with  surface operators  \cite{Wyllard:2010rp,Wyllard:2010vi,Kanno:2011fwx}. When the gauge group is SU(N), the surface operator is labelled by the partition of $N$. In 2d side, it corresponds to $\mathfrak{su}(2)$ embedding of Drinfeld-Sokolov reduction. For the partition $[N-1,1]$, the surface operator is called simple and the corresponding W-algebra was studied by Feigin and Semikhatov in \cite{Feigin:2004wb}.  This algebra has the two currents $e_N(z),f_N(z)$ with spin $\frac{N}{2}$ and all of the other currents with spin $1,2\cdots N-1$ can be obtained by taking the OPE of $e_N(z)$ and $f_N(w)$. 

In this paper, we construct quantum deformation of Feigin-Semikhatov's W-algebras by lifting the Gaiotto-Rap\v{c}\'{a}k's construction to the q-deformed case.  We implement it by gluing two quantum toroidal algebras of $\mathfrak{gl}_1$. In the Gaiotto-Rap\v{c}\'{a}k's framework,  the currents $e_N(z),f_N(z)$ are given by the product of the primary operators for the Y-algebras. For the  purpose  of q-deformation, we need to deform them into the appropriate vertex operators. However, that is difficult to deal with directly because the  definition for q-analogue of a primary field is still vague. Then we first analyze the case of $N=2$ which corresponds to the known algebra $U_q(\widehat{\mathfrak{s}l}_2)$. It is known that $U_q(\widehat{\mathfrak{sl}}_2)$ can be embedded in the quantum toroidal algebra of $\mathfrak{gl}_2$ \cite{Feigin_evaluation}  and how it contains two quantum toroidal algebras of $\mathfrak{gl}_1$ as subalgebras has been already studied in \cite{Feigin:2013fga}. By using these results, we obtain the deformed vertex operators. By extending them to the cases for general $N$, we construct quantum deformation of Feigin-Semikhatov's W-algebras. We also compare the norm of the Whittaker states with the 5d instanton partition function under the existence of a simple surface operator. 

This paper is organized as follows. In section 2, we provide a review of the quantum toroidal algebra of $\mathfrak{gl}_1$ and Gaiotto-Rap\v{c}\'{a}k's VOA. In section 3, we first review the property of the quantum toroidal algebra of $\mathfrak{gl}_2$. Then we decompose $U_q(\widehat{\mathfrak{gl}}_2)$ into the two quantum toroidal algebras of $\mathfrak{gl}_1$, from which we obtain their vertex operators. As an application of the vertex operator for q-$W(\widehat{\mathfrak{gl}}_1)$, we demonstrate that the gluing construction reproduces  the known  representation for the quantum toroidal algebra of $\mathfrak{gl}_2$. In section 4, we extend them to the cases for general $N$ by requiring the commutativity with the screening charges. For the part which cannot be determined from the screening charges, we fix it by demanding that the quadratic relations has the desired property. For the case of $N=3$ which corresponds to deformed Bershadsky-Polyakov algebra \cite{Bershadsky:1990bg,Polyakov:1990}, we write down the quadratic relation in detail. In section 5, we compare the norm of the Whittaker states for the deformed Bershadsky-Polyakov algebra with the 5d SU(3) instanton partition function under the existence of a simple surface operator.

\section{W-algebras associated with $\mathfrak{gl}_{n|m}$ and quantum toroidal algebra of $\mathfrak{gl}_1$}
\label{sec:review}
In \cite{bershtein2018plane}, a new kind of quantum W-algebras was defined as the commutant of the screening charges associated with $\mathfrak{gl}_{n|m}$ root system.
These screening charges appeared in the study of Fock representation of quantum toroidal algebra of $\mathfrak{gl}_1$. On the other hand, 
the same (but undeformed) W-algebras were constructed in a totally different way in \cite{Gaiotto:2017euk}. They arise in the system of the brane junction in type I\hspace{-.1em}IB string theory. 
In this section, we review their definitions  and some properties. 

\subsection{Quantum toroidal algebra of $\mathfrak{gl}_1$}
In this subsection, we follow the convention in \cite{bershtein2018plane}. The quantum toroidal algebra $\mathcal{E}_1(q_1,q_2,q_3)$ of $\mathfrak{gl}_1$ is generated by the Drinfeld currents,
\begin{align}
E(z)=\sum_{m \in \mathbb{Z}}  E_mz^{-m},\;\; F(z)=\sum_{m \in \mathbb{Z}}  F_mz^{-m},\;\;  K^\pm(z)=(C^\perp)^{\pm 1} \exp\left(\sum_{r>0} \mp\frac{\kappa_r}{ r} H_{\pm r} z^{\mp r} \right),
\end{align}
and the centers $C,C^{\perp}$. Here, we set 
\ba
\kappa_r=\prod_{i=1}^3(q_i^{r/2}-q_i^{-r/2})=\sum_{i=1}^3(q_i^{r}-q_i^{-r}).
\ea
The parameters $q_i=\mathrm{e}^{\epsilon_i}\ (i=1,2,3)$ are not independent due to the condition  $\epsilon_1+\epsilon_2+\epsilon_3=0$. The defining relation is given as follows:
\bq
\begin{split}
&g(z,w)E(z)E(w)+g(w,z)E(w)E(z)=0, \qquad\quad g(w,z)F(z)F(w)+g(z,w)F(w)F(z)=0,\\
&K^\pm(z)K^\pm(w) = K^\pm(w)K^\pm (z), 
\quad \qquad
\frac{g(C^{-1}z,w)}{g(C z,w)}K^-(z)K^+ (w) 
=
\frac{g(w,C^{-1}z)}{g(w,C z)}K^+(w)K^-(z),
\\
&g(z,w)K^\pm(C^{(-1\mp1)/2}z)E(w)
+g(w,z)E(w)K^\pm(C^{(-1\mp1) /2}z)=0,
\\
&g(w,z)K^\pm(C^{(-1\pm1)/2}z)
F(w)+g(z,w)F(w)K^\pm(C^{(-1\pm1)/2}z)=0\,,
\\
&[E(z),F(w)]=\frac{1}{\kappa_1}
(\delta\bigl(\frac{Cw}{z}\bigr)K^+(w)
-\delta\bigl(\frac{Cz}{w}\bigr)K^-(z)),
\end{split}
\eq
with some Serre relations.
Here, we set
\begin{align}
 g(z,w)=\prod_{i=1}^3 (z-q_iw),\quad
 \delta(z)=\sum_{m \in \mathbb{Z}}z^m.
\end{align}
Some of the above relations can be also written as follows:
\bq
\begin{split}
&[H_r,H_s]=\delta_{r+s,0}r\frac{C^r-C^{-r}}{\kappa_r},\\
&[H_r,E(z)]=-C^{(-r-|r|)/2}E(z)z^r,\\
&[H_r,F(z)]=C^{(-r+|r|)/2}F(z)z^r.
\end{split}
\eq
For later convenience, we introduce the current $t(z)$ \cite{FHSSY:2010}
\bq
\label{eq:tnewcurrent}
\begin{split}
&t(z)=\alpha(z)E(z)\beta(z),\\
&\alpha(z)=\mathrm{exp}\bigl(\sum_{r=1}^{\infty}\frac{-\kappa_r}{r(1-C^{2r})}H_{-r}z^r\bigr),\\
&\beta(z)=\mathrm{exp}\bigl(\sum_{r=1}^{\infty}\frac{-C^{-r}\kappa_r}{r(1-C^{-2r})}H_rz^{-r}\bigr),
\end{split}
\eq
which commutes with Heisenberg subalgebra $H_r$.
The algebra $\mathcal{E}_1$ is equipped with the coproduct,
\begin{equation}
\begin{aligned}
&\Delta(H_r)=H_r\otimes 1+C^{-r}\otimes H_r,\quad 
\Delta(H_{-r})=H_{-r}\otimes C^{r}+1\otimes H_{-r}, \quad r>0 
\\
&\Delta(E(z))=E\left(C_2^{-1}z\right)\otimes K^+\left(C_2^{-1}z\right)+ 1\otimes E\left(z\right),\\
&\Delta(F(z))=F\left(z\right)\otimes 1 + K^-\left(C_1^{-1}z\right)\otimes F(C_1^{-1}z),\\
&\Delta(X)=X\otimes X,\;\; \text{for $X= C, C^\perp$},
\end{aligned} \label{eq:coprod}
\end{equation} 
where $C_1 =C\otimes 1$, $C_2 =1\otimes C$. The first line implies 
\bq
\Delta K^+(z)=K^+(z)\otimes K^+(C_1z),\quad \Delta K^-(C_2z)=K^-(z)\otimes K^-(z).
\eq

The algebra $\mathcal{E}_1$ has the MacMahone module \cite{Feigin2012quantum} where the bases are labelled by a plane partition. In this representation,  the centers are set to $C=1$ and $C^{\perp}=c$. The currents $K^{\pm}(z)$ are diagonalized and the eigenvalue for the highest weight state $|{\rm hw}\rangle$ is given by
\begin{equation}\label{eq:Macweight}
K^{\pm}(z)|{\rm hw}\rangle=c\frac{1-c^{-2}v/z}{1-v/z}|{\rm hw}\rangle.
\end{equation}
The MacMahone module is irreducible for generic parameters, but, if the condition $c=q_i^{\frac{m}{2}}q_j^{\frac{n}{2}}$ $(i\neq j, n,m\in\mathbb{Z}_{\geq0})$ is satisfied, the states  containing  the box at  $x_i=m+1,x_j=n+1,x_k=1$, ($k\neq i,j$)
become singular\footnote{We set the coordinate of the box at the origin to $(x_1,x_2,x_3)=(1,1,1)$.}.  That implies the irreducible module is described by a plane partition with a "pit"  \cite{bershtein2018plane}. 
One can also introduce an asymptotic Young diagram to each axis, but it must be compatible with the pit condition. Following \cite{bershtein2018plane}, we denote by $\mathcal{M}_{\mu,\nu,\lambda}(v,c)$ the MacMahon module with the  asymptotic Young diagrams $(\mu,\nu,\lambda)$.

In the case of  $c=q_i^{\frac{1}{2}}$, the  MacMahone representation reduces to the Young  diagram representation  which is known as the vertical representation $V^{(i)}_v$ in the literature. One can reconstruct the MacMahone module with $c=q_i^{\frac{m}{2}}q_j^{\frac{n}{2}}$  from the tensor product $V^{(i)}_{v_1}\otimes\cdots\otimes  V^{(i)}_{v_m}\otimes V^{(j)}_{v_{m+1}}\otimes\cdots \otimes V^{(j)}_{v_{n+m}}$, where
\bq
v_l=\begin{cases}
vq_i^{1-l}\quad(l\leq m)\\
vq_i^{1-m}q_j^{m-l}\quad(m<l\leq n).
\end{cases}
\eq
The representation parameters are determined so that the condition (\ref{eq:Macweight}) will hold. 
We note that the order of the tensor product can be freely changed, which is assured by the existence of the universal R-matrix. In the similar way, one can determine the parameters when the asymptotic Young diagram is inserted.
For generic parameters, there are no constraints on the shape of the Young diagrams. In that sense, the MacMahon module corresponds to the degenerate module.

There is another important representation realized by free bosons. We introduce three types of free boson oscillators $a_n$, $n\in\mathbb{Z}$ with the relations\footnote{In this notation, we do not put the subscript $i$ on $a_r$ because one can read it off  from the representation space.}
\begin{equation}
[a_r,a_s]=r\frac{(q_i^{r/2}-q_i^{-r/2})^3}{-\kappa_{r}}\delta_{r+s,0}\qquad i=1,2,3.
\label{eq:ar}
\end{equation}
For later use, we also introduce an operator $Q$ and the parameter $u$ with the relation
\ba
[a_n,Q]=\epsilon_i\delta_{n,0},\quad u=e^{a_0}.
\ea
For each $i$, we have the following free boson representation: 
\bq
\label{eq:horizontalrep}
\begin{split}
\rho_u^{(i)}(E(z))&=\frac{u(1-q_i)}{\kappa_1}\exp\left(\sum_{r=1}^\infty\frac{q_i^{-r/2}\kappa_r}{r(q_i^{r/2}-q_i^{-r/2})^2}a_{-r}z^r\right)\exp\left(\sum_{r=1}^\infty\frac{\kappa_r}{r(q_i^{r/2}-q_i^{-r/2})^2}a_{r}z^{-r}\right), \\
\rho_u^{(i)}(F(z))&=\frac{u^{-1}(1-q_i^{-1})}{\kappa_1}\exp\left(\sum_{r=1}^\infty\frac{-\kappa_r}{r(q_i^{r/2}-q_i^{-r/2})^2}a_{-r}z^r\right)\exp\left(\sum_{r=1}^\infty\frac{-q_i^{r/2}\kappa_r}{r(q_i^{r/2}-q_i^{-r/2})^2}a_{r}z^{-r}\right), \\
\rho_u^{(i)}(H_r)&=\frac{a_r}{q_i^{r/2}-q_i^{-r/2}}, \qquad \rho_u^{(i)}(C^\perp)=1,\qquad \rho_u^{(i)}(C)=q_i^{1/2}.
\end{split}
\eq
We denote the representation space by $\mathcal{F}^{(i)}_u$, or simply by $\mathcal{F}^{(i)}$ if it does  not cause confusion. 

Finally, we mention  the automorphism which changes the centers as $C^{\perp}\to C$, $C\to{(C^{\perp})}^{-1}$. Under the automorphism, the vertical representation $V_u^{(i)}$  transforms into the free boson representation $\mathcal{F}_u^{(i)}$.  That implies the MacMahon representation with $c=q_i^{\frac{m}{2}}q_j^{\frac{n}{2}}$ is isomorphic to the tensor product of free boson representations $\underbrace{\mathcal{F}^{(1)}_{u_1}\otimes\cdots\otimes\mathcal{F}^{(1)}_{u_m}}_{m}\otimes\underbrace{\mathcal{F}^{(2)}_{u_{m+1}}\otimes\cdots\otimes\mathcal{F}^{(2)}_{u_{m+n}}}_{n}$, where the order of the Fock spaces can be changed again.  In this representation, the Drinfeld currents act on the Fock space as vertex operators. They can be characterized by the screening charges introduced in \cite{bershtein2018plane}. Let us first see the simple case  $\mathcal{F}^{(i)}_{u_1}\otimes\mathcal{F}^{(i)}_{u_2}$ which is known to realize q-Virasoro and Heisenberg algebra. It is widely known that there are two screening charges,
\begin{equation}\label{eq:ScrCur+-}
\begin{aligned}
S_{\pm}^{ii}&=\oint S_{\pm}^{ii}(z)\mathrm{d}z,\\
S_+^{ii}(z)&=e^{\frac{\epsilon_{i+1}}{\epsilon_i}Q_{12}}z^{\frac{\epsilon_{i+1}}{\epsilon_i}a_{12}+\frac{\epsilon_i}{\epsilon_{i-1}}}\exp\left(\sum_{r=1}^\infty\frac{-(q_{i+1}^{r/2}{-}q_{i+1}^{-r/2})}{r(q_i^{r/2}{-}q_i^{-r/2})}b_{-r}z^r\right)\exp\left(\sum_{r=1}^\infty\frac{(q_{i+1}^{r/2}{-}q_{i+1}^{-r/2})}{r(q_i^{r/2}{-}q_{i}^{-r/2})} b_{r}z^{-r}\right),\\
S_-^{ii}(z)&=e^{\frac{\epsilon_{i-1}}{\epsilon_i}Q_{12}}z^{\frac{\epsilon_{i-1}}{\epsilon_i}a_{12}+\frac{\epsilon_i}{\epsilon_{i+1}}}\exp\left(\sum_{r=1}^\infty\frac{-(q_{i-1}^{r/2}{-}q_{i-1}^{-r/2})}{r(q_i^{r/2}{-}q_i^{-r/2})}b_{-r}z^r\right)\exp\left(\sum_{r=1}^\infty\frac{(q_{i-1}^{r/2}{-}q_{i-1}^{-r/2})}{r(q_i^{r/2}{-}q_{i}^{-r/2})} b_{r}z^{-r}\right),
\end{aligned}
\end{equation}
where
\bq
\begin{split}
&b_{-r}=q_i^{-r} (a_{-r}\otimes 1)-q_i^{-r/2} (1\otimes a_{-r}), \quad 
b_r=q_i^{r/2} (a_r\otimes 1)-q_i^{r} (1\otimes a_r), \quad r>0,\\
&Q_{12}=Q\otimes 1-1\otimes Q,\quad a_{12}=a_0\otimes 1-1\otimes a_0.
\end{split}
\eq
The next simple case is  $\mathcal{F}^{(i)}_{u_1}\otimes\mathcal{F}^{(j)}_{u_2}$ ($i\neq j$).  The screening charge is given by
\begin{equation}\label{eq:ScrOpe12}
\begin{split}
&S^{ij}=\oint S^{ij}(z)\mathrm{d}z,\\
&S^{ij}(z)=e^{Q'_{12}}z^{a'_{12}+\frac{\epsilon_j}{-\epsilon_i-\epsilon_j}}\exp\left(\sum_{r=1}^\infty \frac{1}{-r} c_{-r}z^r\right)\exp\left(\sum_{r=1}^\infty \frac{1}{r} c_{r}z^{-r}\right),
\end{split}
\end{equation}
where
\begin{equation}
\begin{split}
& c_{-r}=\frac{q_1^{-r}(q_2^{r/2}-q_2^{-r/2})}{q_1^{r/2}-q_1^{-r/2}}(a_{-r}\otimes 1)-\frac{q_1^{-r/2}(q_1^{r/2}-q_1^{-r/2})}{q_2^{r/2}-q_2^{-r/2}}(1\otimes a_{-r}), 
\\
& c_r=\frac{q_1^{r/2}(q_2^{r/2}-q_2^{-r/2})}{q_1^{r/2}-q_1^{-r/2}}(a_r\otimes 1)-\frac{q_3^{-r/2}(q_1^{r/2}-q_1^{-r/2})}{q_2^{r/2}-q_2^{-r/2}}(1\otimes a_r), 
\quad r>0,\\
&Q'_{12}=\frac{\epsilon_j}{\epsilon_i} Q\otimes1-\frac{\epsilon_i}{\epsilon_j} 1\otimes Q,\quad a'_{12}=\frac{\epsilon_j}{\epsilon_i} a_0\otimes1-\frac{\epsilon_i}{\epsilon_j} 1\otimes a_0.
\end{split}
\end{equation}
One can check it satisfies a fermionic relation $S^{ij}(z)S^{ij}(w)=-S^{ij}(w)S^{ij}(z)$.
For the  generic case $\mathcal{F}^{(i_1)}\otimes\mathcal{F}^{(i_2)}\otimes\cdots\otimes\mathcal{F}^{(i_n)}$,
 the screening currents  are given by the union of those for the neighboring Fock spaces $\mathcal{F}^{(i_l)}\otimes\mathcal{F}^{(i_{l+1})}$ $(l=1,\cdots n-1)$. For  $i_1=i_2=\cdots=i_n$, it is known that the screening charges have the structure of $\mathfrak{sl}_n$ root system and give the quantum $W_n$ algebra. For $(\mathcal{F}^{(i)})^{\otimes n}\otimes(\mathcal{F}^{(j)})^{\otimes m}$ ($i\neq j$), the screening charges have the structure similar with the root system of $\mathfrak{sl}_{n|m}$. The corresponding W-algebras have not been well studied so far and the authors of \cite{bershtein2018plane} referred to them as q-$W(\widehat{\mathfrak{gl}}_{n|m})$. The reason why it is not q-$W(\widehat{\mathfrak{sl}}_{n|m})$ but q-$W(\widehat{\mathfrak{gl}}_{n|m})$ is that there is an additional Heisenberg algebra. 

\subsection{Gaiotto-Rap\v{c}\'{a}k's VOA and its extension}
In \cite{Gaiotto:2017euk}, a family of  W-algebras $Y_{L,M,N}[\Psi]$ was constructed from the system of the 5-brane junction with several D3-branes inserted:
\begin{figure}[H]
\begin{center}
\begin{tikzpicture}
\footnotesize
\node (D5) at (1.5,0) {D5 };
\node (NS5) at (0,1.5) {NS5 };
\node (dionic5) at (-1.1,-1.1) { };
\draw (0,0)--(D5);
\draw (0,0)--(NS5);
\draw (0,0)--(dionic5);
\node (D3L) at (-0.8,0.2) {L};
\node (D3M) at (0.3,-0.7) {M};
\node (D3N) at (0.6,0.6) {N};
\end{tikzpicture}
\caption{The nonnegative integers $L,M,N$ indicate the number of the D3-branes. The 5-branes and the D3-branes meet at the three-dimensional intersections. These three intersections have the common two-dimensional boundary at the  junction, where $Y_{L,M,N}[\Psi]$ appears. }
\end{center}
\end{figure}
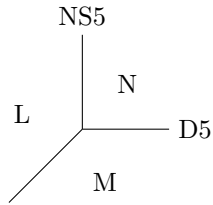
From the viewpoint of D3-branes, we can see the system as $\mathcal{N}=4$ U(L), U(M) and U(N) SYM with the 3d interfaces. By considering twisted $\mathcal{N}=4$ SYM with a coupling constant $\Psi\in\mathbb{C}$, the authors of \cite{Gaiotto:2017euk} obtained the W-algebras $Y_{L,M,N}[\Psi]$ to which they referred as Y-algebra. The explicit form is  given as follows:
\begin{equation}
\label{eq:defYalg}
Y_{L,M,N}[\Psi]=\begin{cases}
\frac{W_{N-M,1,\cdots,1}[\widehat{U}(N|L)_{\Psi}]}{\widehat{U}(M|L)_{\Psi-1}}\quad(N>M)\\
\quad\frac{\widehat{U}(N|L)_{\Psi}\otimes {\rm Sb}^{U(N|L)}}{\widehat{U}(N|L)_{\Psi-1}}\qquad(N=M)\\
\frac{W_{M-N,1,\cdots,1}[\widehat{U}(M|L)_{-\Psi+1}]}{\widehat{U}(N|L)_{-\Psi}}\quad(N<M).
\end{cases}
\end{equation}
Here, we need to explain the notation. The symbol $W_{N_1,N_2\cdots N_n}$ denotes  Drinfeld-Sokolov reduction with $\mathfrak{su}(2)$ embedding labelled by the partition $[N_1,N_2,\cdots ,N_n]$ of $N=N_1+\cdots+N_n$. The level $\Psi$  of $\widehat{U}(N|L)$ is defined so that the subalgebra $\widehat{SU}(N|L)$ will have the level $k=\Psi+L-N$. The matter fields ${\rm Sb}^{U(N|L)}$ consist of $N$ symplectic bosons and $L$ symplectic fermions. They form  $\widehat{U}(N|L)$ and the level of its subalgebra $\widehat{SU}(N|L)$ is $-1$\footnote{The level of $\widehat{SU}(0|L)$ in this notation is different from the standard one by minus sign.}. 

One can also see Y-algebra as a truncation of $W_{1+\infty}$ or affine Yangian of $\mathfrak{gl}_1$\cite{Prochazka:2017qum}\footnote{The claim is subtle when $L,M,N>0$. In this paper, we only consider the cases where one of $L,M$ and $N$ is zero. }. This claim is based on the observation that the vacuum character of $Y_{L,M,N}[\Psi]$ is the same as that of a plane partition with a pit at ($L+1, M+1, N+1$). 
The relation between the  parameters is given by
\ba
\Psi=-\frac{\epsilon_2}{\epsilon_1}.
\ea

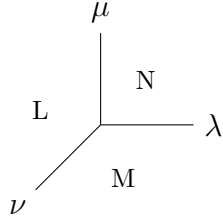
\begin{figure}[H]
\begin{center}
\begin{tikzpicture}
\node (D5) at (1.5,0) {$\lambda$};
\node (NS5) at (0,1.5) {$\mu$};
\node (dionic5) at (-1.1,-1.1) {$\nu$};
\draw (0,0)--(D5);
\draw (0,0)--(NS5);
\draw (0,0)--(dionic5);
\footnotesize
\node (D3L) at (-0.8,0.2) {L};
\node (D3M) at (0.3,-0.7) {M};
\node (D3N) at (0.6,0.6) {N};
\normalsize
\end{tikzpicture}
\caption{The line operators are denoted by their weights  in this figure. When one of $L,M$ or $N$ is zero, it corresponds to the MacMahon module $\mathcal{M}_{\lambda,\mu,\nu}$ with a pit at $(L+1,M+1,N+1)$.\label{fig:asympYoung}}
\end{center}
\end{figure}

The module of Y-algebra is realized by some operators at the junction.  The characteristic one comes from the line operator introduced in the three-dimensional intersections with one of its end points  attached to the junction. The interfaces between, for example, $L$ D3-branes and $M$ D3-branes can be described by U(M$|$L) CS theory and the line operator is labelled by the weight of the gauge group.
In terms of a plane partition, we can interpret it as an asymptotic Young diagram. When we insert the line operators as in Figure \ref{fig:asympYoung},  they give the MacMahon module $\mathcal{M}_{\lambda,\mu,\nu}$.  We note that the shape of the Young diagram is restricted due to the pit and it indeed gives the representation of super Lie group.

We  explain the relation between $Y_{0,M,N}$ and $W(\widehat{\mathfrak{gl}}_{N|M})$ in more detail.  
The simplest case is $Y_{0,0,N}$, which gives $W_N$ algebra and Heisenberg algebra according to (\ref{eq:defYalg}). On the other hand, the MacMahon module with a pit at $(1,1,N+1)$ is equivalent to $(\mathcal{F}^{(3)})^{\otimes n}$, which gives the same algebra. The asymptotic Young diagrams give the completely degenerate module (see \cite{Harada:2018bkb}). The next example is $Y_{0,1,2}$, which gives $\frac{\widehat{U}(2)}{\widehat{U}(1)}$  known as SU(2) parafermion with an extra Heisenberg algebra.  In terms of the quantum toroidal algebra, it should correspond to the vertex operator acting on  the Fock space $\mathcal{F}^{(2)}\otimes(\mathcal{F}^{(3)})^{\otimes 2}$. When we fix the order of the Fock spaces to $\mathcal{F}^{(3)}\otimes\mathcal{F}^{(2)}\otimes\mathcal{F}^{(3)}$, there are two fermionic screening charges. As we will explain,  they are indeed those of SU(2) parafermion. 

One can extend Y-algebra to a wider family of W-algebras by gluing trivalent vertices \cite{Prochazka:2017qum}. In this paper, we mainly focus on the following diagram:
\begin{figure}[H]
\begin{center}
\begin{tikzpicture}
\draw (2,0)--(0,0);
\draw (0,1.5)--(0,0)--(-1.3,-1.3)--(-3.3,-2.4);
\draw (-1.3,-1.3)--(2,-1.3);
\node (2) at (1,0.8) {$N$};
\node (1) at (0.8,-0.8) {1};
\node (01) at (-1.8,-0.2) {0};
\node (02) at (0,-2) {0};
\end{tikzpicture}
\end{center}
\caption{The diagram for Feigin-Semikhatov's W-algebra $ W^{(2)} _N$ \label{fig:FS}}
\end{figure}
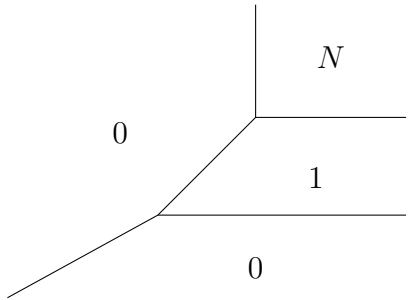
\noindent We first need to comment on the trivalent vertex with different slope.  It can be obtained from the standard one through  the  I\hspace{-.1em}IB S-duality. When the transformation matrix is given by  $M=\left(\begin{array}{c}p\ q\\r\ s\end{array}\right)\in SL(2,\mathbb{Z})$, the D5-brane and NS5-brane are transformed into  $\left(\begin{array}{c}p\\r\end{array}\right)$-brane and $\left(\begin{array}{c}q\\s\end{array}\right)$-brane, respectively. At the same time, the coupling constant changes as $\Psi\to\frac{p\Psi+q}{r\Psi+s}$. In terms of the affine Yangian's parameters, the transformation law is expressed as
\ba
(\epsilon_1,\epsilon_2)\to(\epsilon_1,\epsilon_2)M^{-1}= (s\epsilon_1-r\epsilon_2,-q\epsilon_1+p\epsilon_2). 
\ea
In the q-deformed case, it is rewritten as 
\ba
\label{eq:parameterSdual}
(q_1,q_2)\to(q_1^sq_2^{-r},q_1^{-q}q_2^p).
\ea

According to the BRST procedure, the  algebra associated with figure \ref{fig:FS} is $W_{N-1,1}[\widehat{U}(N)_\Psi]$. It gives affine Kac-Moody algebra $\widehat{U}(2)_\Psi$ for $N=2$ and Bershadsky-Polyakov algebra \cite{Bershadsky:1990bg,Polyakov:1990} for $N=3$. In \cite{Feigin:2004wb}, Feigin and Semikhatov studied the algebra from the two perspectives: screening charges and coset construction. Our concern is on the first one and we will discuss its q-deformation \footnote{We mention that the Y-algebra  provides also the second description. In figure \ref{fig:FS}, the upper corner algebra is $Y_{0,1,N}[\Psi]$, which is equivalent under the I\hspace{-.1em}IB S-duality to  $Y_{N,0,1}[\Psi]=\frac{\widehat{U}(N|1)_{-1+\frac{1}{\Psi}}}{\widehat{U}(N)_{\Psi}}$. This is exactly the coset construction proposed in \cite{Feigin:2004wb}. This duality has been proved in \cite{Creutzig:2020a,Creutzig:2020b}. We also mention that the level-rank duality between the minimal models of $W_M$ and $W_{N-1,1}[\widehat{U}(N)]$ discussed in \cite{Feigin:2004wb} can be described in terms of a plane partition \cite{Harada:2018bkb}.}.
 
One may also consider that  the system consists of $Y_{0,1,N}[\Psi]$, $Y_{0,0,1}[\Psi-1]$ and the  line operators connecting the two junctions. As we have seen, their module is described by two plane partitions.  
The line operators can be interpreted as the asymptotic Young diagrams connecting the two plane partitions.
Then the entire representation space is given by $\oplus_{\lambda\in\mathbb{Z}}\mathcal{M}_{\varnothing,\varnothing,\lambda}\otimes\mathcal{M}_{\varnothing,\lambda,\varnothing}$. We note that  Young diagrams are usually labelled by positive integers, but it was claimed in \cite{Prochazka:2017qum} that negative weight diagrams should be also summed up. While each Y-algebra acts on a single plane partition, we have the other elements which change the asymptotic Young diagram. They correspond to the currents $e_N(z),f_N(z)$  of $W_{N-1,1}[\widehat{U}(N)_\Psi]$ with spin $\frac{N}{2}$. In the following, we sometimes refer to them as  the gluing fields. In terms of free boson realization, they behave as the vertex operators which change the zero mode of the Fock space.

The above description is familiar for $N=2$; it is known that $\widehat{SU}(2)$ can be realized by SU(2) parafermion $\psi(z),\psi^{\dagger}(z)$ and U(1) boson $\phi(z)$ as follows,
\ba
\label{eq:parafermiboson}
J^+(z)=\psi(z)e^{\phi(z)},\quad J^-(z)=\psi^{\dagger}(z)e^{-\phi(z)},\quad J^3(z)=\partial\phi(z).
\ea
In terms of the gluing construction, the currents $J^{\pm}(z)$ correspond to the line operators with $U(1)$ weight $\pm 1$ and are expressed as the product of the primary fields for $Y_{0,1,2}[\Psi]$ and $Y_{0,0,1}[\Psi-1]$\footnote{Precisely speaking, the U(1) boson in (\ref{eq:parafermiboson}) is different from that of $Y_{0,0,1}$. Because the Y-algebra contains  Heisenberg algebra, the primary field of $Y_{0,1,2}$ is given as the product of  $\psi(z)$ (or $\bar{\psi}(z)$) and an extra vertex operator $e^{\tilde\phi(z)}$. For $Y_{0,0,1}$, the primary field is just $e^{\overline\phi(z)}$.  Then we find $\phi(z)=\tilde\phi(z)+\overline\phi(z)$. Another combination of $\tilde\phi(z)$ and $\overline\phi(z)$  commutes with all of the algebra and gives U(1) factor.}. In this construction, the non-localness of the parafermion are resolved thanks to the additional U(1) boson. The same is true for the other $N$ and the conformal dimension of the gluing fields are (half-)integer $\frac{N}{2}$.

It is natural to expect that the algebra associated with the brane-web in figure \ref{fig:FS} is also a truncation of some $W_{\infty}$-like algebra or affine Yangian. We explain this point along with \cite{Prochazka:2017qum,Rapcak:2019abg}.  In general, we can insert D3-branes in the four places as follows:
\begin{figure}[H]
\begin{center}
\begin{tikzpicture}
\draw (2,0)--(0,0);
\draw (0,1.5)--(0,0)--(-1.3,-1.3)--(-3.3,-2.4);
\draw (-1.3,-1.3)--(2,-1.3);
\node (2) at (1,0.8) {L};
\node (1) at (0.8,-0.8) {M};
\node (01) at (-1.8,-0.2) {K};
\node (02) at (0,-2) {N};
\end{tikzpicture}
\end{center}
\caption{The diagram where D3-branes are inserted in the four places.  \label{fig:KLMN}}
\end{figure}
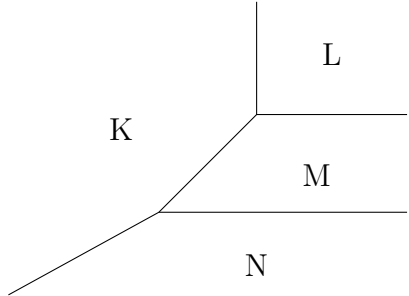
The conformal dimension of the gluing field with fundamental or anti-fundamental weight is $h=1+\frac{L+N-2M}{2}$. That implies the algebras with different values of  $\rho=\frac{L+N-2M}{2}$ should be truncations of  different  $W_{\infty}$-like algebras. Let us consider  the $\rho=0$ case. For $L=M=N=0$, the corresponding algebra is associated with  AGT on $\mathbb{C}^2/\mathbb{Z}_2$ \cite{Nishioka:2011jk}. The relation to AGT becomes more explicit through the string duality, which leads to the system in which  D4-branes stack on the divisor inside the toric Calabi-Yau manifold under the existence of D0-branes \cite{Rapcak:2018nsl,Rapcak:2019abg}. For $L=2,M=1,K=N=0$, it corresponds to $\widehat{U}(2)$.
These facts imply that they are truncations of affine Yangian of $\mathfrak{gl}_2$\footnote{In this paper, we use the terms  $\mathfrak{gl}_n$ and U(n) interchangeably.}. This observation will play an important role in the next section. For the $\rho\neq0$ case, it  includes the non-principal W-algebras $W_{M+2\rho,M}[\widehat{U}(L)]$. The relation between the deformed W-algebras for  general $\mathfrak{su}(2)$ embedding and the quantum toroidal algebras has been discussed in \cite{Negut:2019agq}. We also mention that the deformation of the W-algebra $W_{N,N,\cdots,N}[\widehat{U}(Nn)]$ has been recently proposed in \cite{Negut:2020} together with the relation to the quantum toroidal algebra of $\mathfrak{gl}_n$.

\section{Decomposition of $U_q(\widehat{\mathfrak{gl}}_2)$  by quantum toroidal algebras}
\label{sec:Decomp}
In this section, we explore whether the gluing construction works even after q-deformation. We study how $U_q(\widehat{\mathfrak{gl}}_2)$ is decomposed into q-$W(\widehat{\mathfrak{gl}}_{2|1})$ and q-$W(\widehat{\mathfrak{gl}}_{1})$. For that purpose, it is better to embed $U_q(\widehat{\mathfrak{gl}}_2)$ into the quantum toroidal algebra of $\mathfrak{gl}_2$. We first review the definition of the  quantum toroidal algebra of $\mathfrak{gl}_2$ and its important properties following \cite{Feigin:2013fga,Feigin:2018moi}.  

\subsection{Quantum toroidal algebra of $\mathfrak{gl}_2$}
The quantum toroidal algebra $\mathcal{E}_2(q_1,q_2,q_3)$ of $\mathfrak{gl}_2$ has two parameters $d,q$. The following parameters are also used,
\begin{align}
& q_1=d q^{-1},\ q_2=q^2,\ q_3=d^{-1}q^{-1}\ .
\end{align}
It has the following generators,
\begin{align}
E_{i,k},\ F_{i,k},\ H_{i,r},\ K_i^{\pm1},\ C'^{\pm 1} \quad  
(i\in \{0,1\},\ k\in\Z,\  
r\in\Z/\{0\}).
\end{align}
By introducing the Drinfeld currents
\begin{align}
E_i(z) =\sum_{k\in \Z}E_{i,k}z^{-k}, \quad 
F_i(z) =\sum_{k\in\Z}F_{i,k}z^{-k}, \quad
K_i^{\pm}(z) = K_i^{\pm 1} \exp(\pm(q-q^{-1})\sum_{r=1}^\infty H_{i,\pm r}z^{\mp r}),
\end{align}
the defining relations are written as follows:
\bq
\begin{split}
&\hspace{40mm}\text{$C'^{\pm 1}$ are central},\\
&\hspace{10mm}K_iE_j(z)K_i^{-1}=q^{a_{i,j}}E_j(z),\quad K_iF_j(z)K_i^{-1}=q^{-a_{i,j}}F_j(z),\\
&\hspace{30mm}K^\pm_i(z)K^\pm_j (w) = K^\pm_j(w)K^\pm_i (z), \\
&\hspace{10mm}\frac{g_{i,j}(C'^{-1}z,w)}{g_{i,j}(C'z,w)}K^-_i(z)K^+_j (w) =
\frac{g_{j,i}(w,C'^{-1}z)}{g_{j,i}(w,C'z)}K^+_j(w)K^-_i (z),\\
&(-1)^{i+j}g_{i,j}(z,w)K_i^\pm(C'^{(-1\mp1)/2}z)E_j(w)+g_{j,i}(w,z)E_j(w)K_i^\pm(C'^{(-1\mp1)/2}z)=0,\\
&(-1)^{i+j}g_{j,i}(w,z)K_i^\pm(C'^{(-1\mp1)/2}z)F_j(w)+g_{i,j}(z,w)F_j(w)K_i^\pm(C'^{(-1\mp1)/2}z)=0\,,\\
&\hspace{10mm}[E_i(z),F_j(w)]=\frac{\delta_{i,j}}{q-q^{-1}}
(\delta\bigl(C'\frac{w}{z}\bigr)K_i^+(w)
-\delta\bigl(C'\frac{z}{w}\bigr)K_i^-(z)),\\
&\hspace{10mm}(-1)^{i+j}g_{i,j}(z,w)E_i(z)E_j(w)+g_{j,i}(w,z)E_j(w)E_i(z)=0, \\
&\hspace{10mm}(-1)^{i+j}g_{j,i}(w,z)F_i(z)F_j(w)+g_{i,j}(z,w)F_j(w)F_i(z)=0,\\
\end{split}
\eq
where 
\begin{equation}
a_{i,j}=
\begin{pmatrix}
2 & -2\\
-2 & 2\\
\end{pmatrix}
	     \label{eq:Cartansl2}
\end{equation}
is the Cartan matrix of $\widehat{\mathfrak{sl}}_2$ and
\begin{align}
g_{i,j}(z,w)=\begin{cases}
	      z-q_2w & (i= j),\\
              (z-q_1w)(z-q_3w)& (i\neq j).
	     \end{cases}
\end{align}
It is possible to rewrite the above relations using $H_{i,r}$ instead of $K^{\pm}_i(z)$ as follows:
\begin{align}
&[H_{i,r},E_j(z)]=a_{i,j}(r) z^r E_j(z)C'^{-(r+|r|)/2}\,,
\\
&[H_{i,r},F_j(z)]=-a_{i,j}(r)z^r F_j(z)C'^{-(r-|r|)/2}\,,
\\
&[H_{i,r},H_{j,s}]=
\delta_{r+s,0}\,a_{i,j}(r)\frac{C'^r-C'^{-r}}{q-q^{-1}},
\end{align}
where $a_{i,i}(r)=[r]_q(q^r+q^{-r})/r$, 
 $a_{i,j}(r)=-[r]_q(d^r+d^{-r})/r$ ($i\neq j$) and $[r]_q=\frac{q^r-q^{-r}}{q-q^{-1}}$.
 
The algebra $\mathcal E_2(q_1,q_2,q_3)$ has the subalgebras $\mathcal E_1^{(1)}(\tilde{q}_1,\tilde{q}_2,\tilde{q}_3)$ and $\mathcal E_1^{(3)}(\hat{q}_1,\hat{q}_2,\hat{q}_3)$ commuting with each other. Their generators and parameters are given, respectively,  as follows \cite{Feigin:2013fga}:
\begin{eqnarray}
\label{eq:gl1sub1}
\text{For $\mathcal E_1^{(1)}$},\quad &E(z)=E_{1|0}^{(1)},\ F(z)=F_{1|0}^{(1)}(z),\ K(z)=K_{1|0}^{\pm\ (1)}(z), C=C', C^{\perp}=K_1K_2,\\
&\tilde{q}_1=q_2,\ \  \tilde{q}_2=q_1^2,\ \ \tilde{q}_3=q_3q_1^{-1}. \label{eq:upper}\\
\ \nonumber\\
\label{eq:gl1sub2}
\text{For $\mathcal E_1^{(3)}$},\quad &E(z)=E_{1|0}^{(3)},\ F(z)=F_{1|0}^{(3)}(z),\ K(z)=K_{1|0}^{\pm\ (3)}(z), C=C', C^{\perp}=K_1K_2,\\
&\hat{q}_1=q_2,\ \ \hat{q}_2=q_1q_3^{-1},\ \ \hat{q}_3=q_3^2\label{eq:lower}.
\end{eqnarray}
Here, we set
\bq
\label{eq:10current}
\begin{split}
&E_{1|0}^{(1)}(z)=\lim_{z'\to z}(1-\frac{z}{z'})(1-\frac{q_3z}{q_1z'})E_1 (q_1z')E_0(z), \\
&F_{1|0}^{(1)}(z)=\lim_{z'\to z}(1-\frac{z'}{z})(1-\frac{q_1z'}{q_3z})F_0(z) F_1 (q_1z'), \\
&K_{1|0}^{\pm,(1)}(z)=K_1^{\pm}(q_1z)K_0^{\pm}(z),\\
&E_{1|0}^{(3)}(z)=\lim_{z'\to z}(1-\frac{z}{z'})(1-\frac{q_1z}{q_3z'})E_1 (q_3z')E_0(z),\\
&F_{1|0}^{(3)}(z)=\lim_{z'\to z}(1-\frac{z'}{z})(1-\frac{q_3z'}{q_1z})F_0(z) F_1 (q_3z'), \\
&K_{1|0}^{\pm,(3)}(z)=K_1^{\pm}(q_3z)K_0^{\pm}(z).
\end{split}
\eq

Another important subalgebra is the quantum affine algebra $U_q(\widehat{\mathfrak{sl}}_2)$. If we set 
\bq
\label{eq:uqsl2sub}
\begin{split}
&x^+(z)
=E_1(\alpha z),\ \ x^-(z)=F_1(\alpha z),\ \ h_r=(C'^{1/2}\alpha^{-1})^rH_{1,r}\ (r\in\mathbb{Z}/0),\ \  q^{h_0}=K_1,\\
&  \sum_{m=0}^\infty \psi_m z^{-m}=q^{h_0}\mathrm{exp}\bigl((q-q^{-1})\sum_{m=1}^\infty
h_m z^{-m}\bigr),\\
 &\sum_{m=0}^\infty \varphi_{-m}
z^{m}=q^{-h_0}\mathrm{exp}\bigl(-(q-q^{-1})\sum_{m=1}^\infty
h_{-m} z^{m}\bigr),
\end{split}
\eq
these elements satisfy the following defining relation of $U_q(\widehat{\mathfrak{sl}}_2)$:
\bq
\begin{split}
&[h_m,h_n]=\delta_{m+n,0}{{[2m]_q[mc]_q}\over{m}},\ [h_m,q^{h_0}]=0,\\
&q^{h_0}x_m^\pm q^{-h_0}=q^{\pm2}x_m^\pm,\\
& [h_m,x_n^\pm]=
\pm \frac{[2m]_q}{m}q^{\mp \frac{|m|c}{2}}x_{m+n}^\pm,\\
&x_{m+1}^\pm x_{n}^\pm -q^{\pm2}x_n^\pm x_{m+1}^\pm=
q^{\pm2} x_{m}^\pm x_{n+1}^\pm -x_{n+1}^\pm x_{m}^\pm,\\
&[x_m^+,x_n^-]={1\over{q-q^{-1}}}(q^{{{(m-n)}\over2}c}\psi_{m+n}-q^{{{(n-m)}\over2}c}
\varphi_{m+n}),
\end{split}
\eq
where $C'=q^c$. The parameter $\alpha$ can take an arbitrary value, but we fix it to $\alpha=q_3^{1/2}q^{-1}$ in the following.

In \cite{Shiraishi:1992nqz,MatsuoA:1992}, the free field realization of $U_q(\widehat{\mathfrak{sl}}_2)$ was constructed. These results were promoted to the representation $\rho$ of $\mathcal{E}_2$ in \cite{Feigin_evaluation,Feigin:2018moi}, which is so-called evaluation homomorphism. In this representation,  the center takes the value $C'=q_3$ and the generators behave as the following vertex operators:
\bq
\label{eq:gl2rep}
\begin{split}
&-(q-q^{-1})\rho(E_1(z))=:e^{u_1^+(z)}:-:e^{u_1^-(z)}:,\quad (q-q^{-1})\rho(F_1(z))=:e^{v_1^+(z)}:-:e^{v_1^-(z)}:,\\
&-(q-q^{-1})\rho(E_0(z))=:e^{\widehat{u}_1^+(z)}:-:e^{\widehat{u}_1^-(z)}:,\quad (q-q^{-1})\rho(F_0(z))=:e^{\widehat{v}_1^+(z)}:-:e^{\widehat{v}_1^-(z)}:,\\
&\rho(H_{1,n})=h_{1,n},\quad \rho(H_{0,n})=\widehat{h}_{1,n}, \quad \rho(K_1)=q^{h_{1,0}},\quad \rho(K_0)=q^{-h_{1,0}},
\end{split}
\eq
where the following free boson fields are introduced
\begin{align}
&u^\pm_1(z)=Q_{u_1}+u_{1,0}\log z
\mp a_{3,0}\log q-\sum_{n\neq0}\frac{u^\pm_{1,n}}{n}z^{-n}\,,
\label{u1n}\\
&v^\pm_1(z)=
-\bigl(Q_{u_1}+u_{1,0}\log z\bigr)
\pm a_{2,0}\log q -\sum_{n\neq0}\frac{v^\pm_{1,n}}{n}z^{-n},\\
&\hat{u}^\pm_1(z)=
-\bigl(Q_{u_1}+u_{1,0}\log z\bigr)
\mp a_{2,0}\log q-\sum_{n\neq0}\frac{\hat{u}^\pm_{1,n}}{n}z^{-n}\,,
\label{uh1}\\
&\hat{v}^\pm_1(z)=Q_{u_1}+u_{1,0}\log z
\pm a_{3,0}\log q
-\sum_{n\neq0}\frac{\hat{v}^\pm_{1,n}}{n}z^{-n}.
\end{align}
We note that the above bosons are not independent and are expressed by four independent bosons. One of them corresponds to $\widehat{\mathfrak{gl}}_1$ and the other three correspond to the bosons used in Wakimoto representation. The concrete expression for these bosons are not necessary in what follows and we omit it. See \cite{Feigin:2018moi} for details.

\subsection{Decomposition of  $U_q(\widehat{\mathfrak{gl}}_2)$}
Let us explore the decomposition of  $\mathcal{E}_2$ into the two $\mathcal{E}_1$s in the case where they satisfy the truncation conditions.  Because the centers of  the subalgebras $\mathcal{E}_1^{(1)}$ and $\mathcal{E}_1^{(3)}$ are the same, their truncation conditions are not independent: the nonnegative integer parameters $K,M,L,K',N,M'$ labelling each truncation condition must satisfy the relation,
\ba
C=\tilde{q}_1^{K}\tilde{q}_2^{M}\tilde{q}_3^{L}=\hat{q}_1^{K'}\hat{q}_2^{N}\hat{q}_3^{M'}.
\ea
One can see from (\ref{eq:upper}) and (\ref{eq:lower}) that it implies
\ba
\label{eq:KLMN}
K=K',\quad M=M', \quad L+N-2M=0
\ea
for  generic  $q_1,q_2,q_3$.
Let us compare these results with Gaiotto-Rap\v{c}\'{a}k's construction. In figure \ref{fig:KLMN}, there are two Y-algebras, $Y_{K,M,L}$ and $Y_{K,N,M}$. This combination matches  the first two relations in (\ref{eq:KLMN}). The last condition also matches the discussion in the last part of section \ref{sec:review}; as we have explained, the $W_{\infty}$-like algebra associated with figure \ref{fig:KLMN} depends on the parameter $\rho=\frac{L+N-2M}{2}$ and it is expected that the affine Yangian of $\mathfrak{gl}_2$ is realized in the $\rho=0$ case. Further, the relation between the parameter of $\mathcal{E}_1^{(1)}$ and that of $\mathcal{E}_1^{(3)}$ is also consistent with the relation  (\ref{eq:parameterSdual}) in the gluing construction.

In the remaining part of this section, we focus on the  case of $C=q_3$ where $\mathcal{E}_2$ truncates to $U_q(\widehat{\mathfrak{gl}}_2)$. 
The subalgebras $\mathcal{E}_1^{(1)}$ and $\mathcal{E}_1^{(3)}$ truncate to q-$W(\widehat{\mathfrak{gl}}_{2|1})$\footnote{The algebra q-$W({\mathfrak{sl}_{2|1}})$ was originally proposed in \cite{Ding:1998si} and further studied in \cite{Kojima:2019ewe}.} and q-$W(\widehat{\mathfrak{gl}}_1)$, respectively. As we have seen, q-$W(\widehat{\mathfrak{gl}}_{2|1})$ (resp. q-$W(\widehat{\mathfrak{gl}}_1)$) can be realized by three bosons (resp. one boson). In our notation, we express the boson for q-$W(\widehat{\mathfrak{gl}}_1)$
with a tilde and the boson for q-$W(\widehat{\mathfrak{gl}}_{2|1})$ acting on $i$-th Fock space with a subscript $(i)$\footnote{One should be careful not to confuse the subscript $(i)$ with that in ({\ref{eq:ar}}). Recall that we do not put the subscript which distinguish the type of the boson.}. In order to understand how the gluing fields of $U_q(\widehat{\mathfrak{gl}}_2)$ decompose into the vertex operators for the two q-W algebras, let us  express the bosons used in (\ref{eq:gl2rep}) by these bosons. 
We can implement that by comparing the two  free boson representations for $\mathcal{E}_1$ : the first one is obtained from  (\ref{eq:10current}) and (\ref{eq:gl2rep})
while the other one is directly obtained from  (\ref{eq:coprod}) and (\ref{eq:horizontalrep}). The result is as follows (see Appendix \ref{app:e1e0} for the detail of the computation):
\ba
\label{eq:upm}
\begin{split}
&u_{1,n}^+=
\begin{cases}
-\frac{(q^n-q^{-n})q_3^n}{q_1^n-q_1^{-n}}a_n^{(2)}-\frac{(q^n-q^{-n})(q_1^n-q_1^{-n})q_3^n}{d^n-d^{-n}}a_n^{(3)}-\frac{(q^n-q^{-n})q^{-n}}{q_3^n-q_3^{-n}}\tilde{a}_n\quad(n>0) \\
-\frac{(q^n-q^{-n})q^{-2n}}{q_1^n-q_1^{-n}}a_n^{(2)}-\frac{(q^n-q^{-n})q_3^nq^{-n}}{q_3^n-q_3^{-n}}\tilde{a}_n\qquad(n<0)
\end{cases}\\
&u_{1,n}^-=
\begin{cases}
-\frac{(q^n-q^{-n})q_1^n}{q_1^n-q_1^{-n}}a_n^{(2)}-\frac{(q^n-q^{-n})q^{-n}}{q_3^n-q_3^{-n}}\tilde{a}_n\quad(n>0) \\
-\frac{(q^n-q^{-n})q_1^{2n}}{q_1^n-q_1^{-n}}a_n^{(2)}+\frac{(q^n-q^{-n})(q_1^n-q_1^{-n})q^{-n}}{d^n-d^{-n}}a_n^{(3)}-\frac{(q^n-q^{-n})q_3^nq^{-n}}{q_3^n-q_3^{-n}}\tilde{a}_n\qquad(n<0)
\end{cases}\\
\end{split}
\ea
\ba
\label{eq:vpm}
\begin{split}
&v_{1,n}^+=
\begin{cases}
-\frac{(q^n-q^{-n})(q_1^n-q_1^{-n})q^{-n}}{d^n-d^{-n}}a_n^{(1)}+\frac{(q^n-q^{-n})q_1^{2n}}{q_1^n-q_1^{-n}}a_n^{(2)}+\frac{(q^n-q^{-n})q^{-n}q_3^n}{q_3^n-q_3^{-n}}\tilde{a}_n\quad(n>0) \\
\frac{(q^n-q^{-n})q_1^n}{q_1^n-q_1^{-n}}a_n^{(2)}+\frac{(q^n-q^{-n})q^{-n}}{q_3^n-q_3^{-n}}\tilde{a}_n\qquad(n<0)
\end{cases}\\
&v_{1,n}^-=
\begin{cases}
\frac{(q^n-q^{-n})q^{-2n}}{q_1^n-q_1^{-n}}a_n^{(2)}+\frac{(q^n-q^{-n})q^{-n}q_3^n}{q_3^n-q_3^{-n}}\tilde{a}_n\quad(n>0) \\
\frac{(q^n-q^{-n})(q_1^n-q_1^{-n})q_3^n}{d^n-d^{-n}}a_n^{(1)}+\frac{(q^n-q^{-n})q_3^n}{q_1^n-q_1^{-n}}a_n^{(2)}+\frac{(q^n-q^{-n})q^{-n}}{q_3^n-q_3^{-n}}\tilde{a}_n\qquad(n<0)
\end{cases}\\
\end{split}
\ea
\footnotesize
\ba
\begin{split}
h_{1,n}=
\begin{cases}
&\hspace{-3mm}\frac{[n]_q}{n}\left(\frac{q_1^n-q_1^{-n}}{d^n-d^{-n}}a_n^{(1)}-d^na_n^{(2)}+\frac{(q_1^n-q_1^{-n})q^n}{d^n-d^{-n}}a_n^{(3)}-\tilde{a}_n\right)\ (n>0)\hspace{30mm}\\
&\hspace{-3mm}\frac{[n]_q}{n}\left(\frac{(q_1^n-q_1^{-n})q^n}{d^n-d^{-n}}a_n^{(1)}-d^na_n^{(2)}+\frac{q_1^n-q_1^{-n}}{d^n-d^{-n}}a_n^{(3)}-\tilde{a}_n\right)\ (n<0)
\end{cases}
\end{split}
\label{eq:h1n}
\ea
\normalsize
\ba
\label{eq:zeromodeQ}
\begin{split}
&Q_{u_1}=Q^{(2)}+\tilde{Q},\quad u_{1,0}=\frac{1}{\epsilon_1-\epsilon_3}(a_0^{(2)}-\tilde{a}_0),\quad h_{1,0}=\frac{1}{\epsilon_1-\epsilon_3}(-\epsilon_3a_0^{(2)}+\epsilon_1\tilde{a}_0)\\
&a_{2,0}=\frac{1}{\epsilon_2}(-a_0^{(1)}+a_0^{(2)}),\quad a_{3,0}=\frac{1}{\epsilon_2}(-a_0^{(2)}+a_0^{(3)}),
\end{split}
\ea
where the commutation relations  are given as follows:
\ba
\begin{split}
&[a_n^{(1)},a_m^{(1)}]=[a_n^{(3)},a_m^{(3)}]=-n\frac{(d^n-d^{-n})^2}{(q_1^n-q_1^{-n})(q^n-q^{-n})}\delta_{n+m,0},\\
&[a_n^{(2)},a_m^{(2)}]=n\frac{(q_1^n-q_1^{-n})^2}{(d^n-d^{-n})(q^n-q^{-n})}\delta_{n+m,0},\quad [\tilde{a}_n,\tilde{a}_m]=-n\frac{(q_3^n-q_3^{-n})^2}{(d^n-d^{-n})(q^n-q^{-n})}\delta_{n+m,0},\\
&[a_n^{(2)},Q^{(2)}]=[\tilde{a}_n,\tilde{Q}]=\epsilon_2\delta_{n,0}.
\end{split}
\label{eq:subboson}
\ea
The other commutators vanish. The zero modes satisfy the condition,
\ba
\epsilon_1u_{1,0}=\frac12(a_0^{(1)}+a_0^{(3)}).
\ea
We note that this condition is compatible in the entire Fock space.
 The vertex operator for q-$W(\widehat{\mathfrak{gl}}_{2|1})$ is realized on the Fock space $\mathcal{F}^{(3)}\otimes\mathcal{F}^{(2)}\otimes\mathcal{F}^{(3)}$  while that for q-$W(\widehat{\mathfrak{gl}}_1)$ is realized on $\tilde{\mathcal{F}}^{(3)}$. The reason why the order of Fock spaces for q-$W(\widehat{\mathfrak{gl}}_{2|1})$ is fixed to $\mathcal{F}^{(3)}\otimes\mathcal{F}^{(2)}\otimes\mathcal{F}^{(3)}$ is that  the Wakimoto representation of $\widehat{\mathfrak{sl}}_2$  can be characterized by two fermionic screening charges\footnote{We note that  there is an additional screening charge which is not an exponential type but dressed by a polynomial.};  for another ordering of the Fock spaces, we only have one fermionic screening charge.
 
It is convenient to introduce the boson $a_n^{\mathfrak{u}(1)}$ associated with Heisenberg subalgebra of q-$W(\widehat{\mathfrak{gl}}_{2|1})$ as follows:
\bq
\label{eq:defanu1}
\frac{a_n^{\mathfrak{u}(1)}}{q_3^n-q_3^{-n}}=\begin{cases}
-\frac{a_n^{(1)}}{d^n-d^{-n}}+\frac{d^na_n^{(2)}}{q_1^n-q_1^{-n}}-\frac{q^na_n^{(3)}}{d^n-d^{-n}}\quad(n>0)\\
-\frac{q^na_n^{(1)}}{d^n-d^{-n}}+\frac{d^na_n^{(2)}}{q_1^n-q_1^{-n}}-\frac{a_n^{(3)}}{d^n-d^{-n}}\quad(n<0).
\end{cases}
\eq
Then, the relation (\ref{eq:h1n}) can be rewritten into the simpler form:
\bq
h_{1,n}=-\frac{[n]_q}{n}J_n,\qquad J_n=\frac{q_1^n-q_1^{-n}}{q_3^n-q_3^{-n}}a_n^{\mathfrak{u}(1)}+\tilde{a}_n.
\eq
The other linear combination of $a_n^{\mathfrak{u}(1)}$ and $\tilde{a}_n$ which commutes with $J_n$ is 
\bq
j_n=a_n^{\mathfrak{u}(1)}+\tilde{a}_n.
\eq
It commutes with all the elements of $U_q(\widehat{\mathfrak{sl}}_2)$ and forms $U_q(\widehat{\mathfrak{gl}}_2)$ together with $U_q(\widehat{\mathfrak{sl}}_2)$. 

We can extract the Heisenberg part from the currents $x^{\pm}(z)$ as follows:
\begin{align}
\label{eq:xpdecomposition}
&x^+(z)=\psi_q(z) :\exp\left(\sum_{n\neq0}\frac{q_3^{-\frac{|n|}{2}}(q^n-q^{-n})}{n(q_3^n-q_3^{-n})}J_nz^{-n}\right):,\\
&x^-(z)=\overline{\psi}_q(z) :\exp\left(-\sum_{n\neq0}\frac{q_3^{\frac{|n|}{2}}(q^n-q^{-n})}{n(q_3^n-q_3^{-n})}J_nz^{-n}\right):.
\label{eq:xmdecomposition}
\end{align}
The vertex operators $\psi_q(z), \overline{\psi}_q(z)$ can be interpreted  as q-deformed SU(2) parafemions.

Finally, we mention the screening currents of the deformed Wakimoto representation. As we have explained,  there are two fermionic screening currents in $\mathcal{F}^{(3)}\otimes\mathcal{F}^{(2)}\otimes\mathcal{F}^{(3)}$. We checked that they are indeed the same as those for $U_q(\widehat{\mathfrak{sl}}_2)$ obtained in \cite{Matsuo:1992va}.

\subsection{The vertex representation of $\mathcal{E}_2$ from the gluing construction}
As an application of the vertex operator for q-$W(\widehat{\mathfrak{gl}}_1)$, we consider the gluing construction for $K=1,L=M=N=0$:
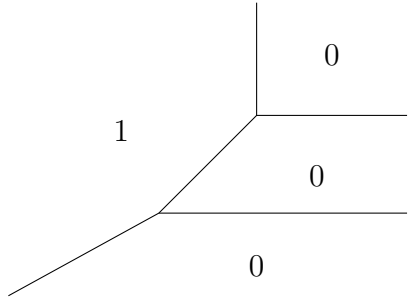
\begin{center}
\begin{tikzpicture}
\draw (2,0)--(0,0);
\draw (0,1.5)--(0,0)--(-1.3,-1.3)--(-3.3,-2.4);
\draw (-1.3,-1.3)--(2,-1.3);
\node (2) at (1,0.8) {$0$};
\node (1) at (0.8,-0.8) {0};
\node (01) at (-1.8,-0.2) {1};
\node (02) at (0,-2) {0};
\end{tikzpicture}
\end{center}
 The branching rule of the module has been already studied in \cite{Feigin:2013fga}, where the vertical representation  ($C=1$) was considered. In this section, we consider the free boson representation. From the truncation condition, we find that the center should be $C'=q$. 
In the previous case, the Drinfeld currents $E_1(z)$, $F_1(z)$ of $\mathcal{E}_2$ are  realized as the product of the vertex operators for q-$W(\widehat{\mathfrak{gl}}_{2|1})$ and q-$W(\widehat{\mathfrak{gl}}_1)$.
In the following, we examine whether the gluing construction works also for the present case. 
We refer to q-$W(\widehat{\mathfrak{gl}}_1)$ associated with the upper (resp. the lower) junction as the upper (resp. the lower) q-W. 

The vertex operators $V^{\pm}(z)$ for q-$W(\widehat{\mathfrak{gl}}_1)$ which we found in the previous subsection are rewritten as
\ba
\label{eq:u1vertex1}
V^{\pm}(z)=\exp\left(\pm(\tilde{Q}'+\tilde{a}_0'\log z-\sum_{n\neq0}q_3^{\mp\frac{|n|}{2}}\frac{\tilde{a}'_n}{n}z^{-n})\right),
\ea
where
\ba
[\tilde{a}_0',\tilde{Q}']=\frac{\epsilon_2}{\epsilon_3-\epsilon_1},\quad [\tilde{a}_n',\tilde{a}_m']=-n\frac{q^n-q^{-n}}{d^n-d^{-n}}\delta_{n+m,0}.
\ea
To apply these vertex operators to the present case, we need slight modification because the parameters of q-$W(\widehat{\mathfrak{gl}}_1)$ are not the same. In the MacMahon language, the vertex operators (\ref{eq:u1vertex1}) correspond to $\mathcal{M}_{\emptyset, \mu,\emptyset}$ with a pit at $(1,1,2)$, where $\mu$ is either the fundamental weight or the anti-fundamental weight. On the other hand, the vertex operators for the lower q-W correspond to $\mathcal{M}_{\emptyset,\mu,\emptyset}$ with a pit at $(2,1,1)$. Then they are obtained  from (\ref{eq:u1vertex1}) by exchanging the parameters $\hat{q}_1$ and $\hat{q}_3$ defined in (\ref{eq:lower}), which leads to
\ba
&&V^{\pm}_l(z)=\exp\left(\pm(Q_l+l_0\log z-\sum_{n\neq0}q^{\mp\frac{|n|}{2}}\frac{l_n}{n}z^{-n})\right),\\
&&[l_0,Q_l]=\frac{2\epsilon_3}{\epsilon_3-\epsilon_1},\quad [l_n,l_m]=-n\frac{q_3^n-q_3^{-n}}{d^n-d^{-n}}\delta_{n+m,0}.
\ea
We can obtain the vertex operators $V^{\pm}_u(z)$ for the upper q-W   in the similar way. 
They correspond to $\mathcal{M}_{\emptyset,\emptyset,\mu}$ with a pit at $(2,1,1)$ and we need to exchange $\hat{q}_2$ and $\hat{q}_3$. We also need to change $\hat{q}_i$ to $\tilde{q}_i$. As a result, we have
\ba
&&V^{\pm}_u(z):=\exp\left(\pm(Q_u+u_0\log z-\sum_{n\neq0}q^{\mp\frac{|n|}{2}}\frac{u_n}{n}z^{-n})\right),\\
&&[u_0,Q_u]=\frac{2\epsilon_1}{\epsilon_1-\epsilon_3},\quad [u_n,u_m]=n\frac{q_1^n-q_1^{-n}}{d^n-d^{-n}}\delta_{n+m,0}.
\ea

If the gluing construction works, the following vertex operators should give the representation of $\mathcal{E}_2$:
\ba
E_1(z)\to V_l^+(z)V_u^+(z),\quad F_1(z)\to V_l^-(z)V_u^-(z).
\ea
We can simplify them as follows:
\ba
V_d^{\pm}(z)\equiv V_l^{\pm}(z)V_u^{\pm}(z)=\exp\left(\pm(Q_d+d_0\log z-\sum_{n\neq0}q^{\mp\frac{|n|}{2}}\frac{d_n}{n}z^{-n})\right),
\ea
where
\ba
[d_0,Q_d]=2,\quad [d_n,d_m]=n(q^n+q^{-n})\delta_{n+m,0}.
\ea
One can check that they obey the quadratic relations of $\mathcal{E}_2$ if we set
\ba
H_{1,n}\to\frac{q^{n/2}-q^{-3n/2}}{q-q^{-1}}\frac{d_n}{n}.
\ea
The other currents $E_0(z),F_0(z),H_{0,n}$ are given in the same way by introducing the boson whose commutators with $d_n$ and itself are proportional to (\ref{eq:Cartansl2}). One can check that they indeed give the $C'=q$ representation of $\mathcal{E}_2$. This type of the vertex representation was studied in \cite{Saito:1996ai}  (see also \cite{Awata:2017lqa}). The above result suggests that the gluing construction works well at least for q-$W(\widehat{\mathfrak{gl}}_1)$.

\section{Quantum deformation of Feigin-Semikhatov's W-algebra}
\label{sec:FS}
In this section, we discuss quantum deformation of Feigin-Semikhatov's W-algebra $\mathcal{W}_N^{(2)}$.  The generating currents $e_N(z), f_N(z)$ of $\mathcal{W}_N^{(2)}$ were constructed as the commutant of a set of screening charges in \cite{Feigin:2004wb}. In this construction, the additional Heisenberg algebra which trivially commutes with the screening charges is also introduced to make   the currents local.  As we have seen, one can construct  $\mathcal{W}_N^{(2)}$ by gluing $Y_{0,1,N}$ and $Y_{0,0,1}$. In the following, we construct its q-deformation by gluing q-$W(\widehat{\mathfrak{gl}}_{N|1})$ and q-$W(\widehat{\mathfrak{gl}}_1)$. In the following,  the deformation of $e_N(z)$ and $f_N(z)$  is denoted by $\mathscr{E}_N(z)$ and $\mathscr{F}_N(z)$. 


\subsection{Quantum deformation of Bershadsky-Polyakov algebra}
Let us first consider the case of $N=3$ which corresponds to  Bershadsky-Polyakov algebra. There are several possibilities for the order of Fock spaces of q-$W(\widehat{\mathfrak{gl}}_{3|1})$. In this section, we consider $\mathcal{F}^{(3)}\otimes\mathcal{F}^{(2)}\otimes\mathcal{F}^{(3)}\otimes\mathcal{F}^{(3)}$. For each pair of neighboring Fock spaces, we have a screening charge. The screening charges  for the first three Fock spaces are the same as  those of q-$W(\widehat{\mathfrak{gl}}_{2|1})$ and we have already studied them in the previous section.  The new one is  $S^{33}_{\pm}$ which acts on the last two Fock spaces. Because we already know that the vertex operators for q-$W(\widehat{\mathfrak{gl}}_{2|1})$ commute with the screening charges except $S^{33}_{\pm}$, it is natural to consider that the vertex operators of q-$W(\widehat{\mathfrak{gl}}_{3|1})$ can be obtained by some modification.
Actually, we do not need any modification for $\mathscr{F}_3(z)$. That is because it acts only on the first two Fock spaces as we can see from (\ref{eq:vpm}) and it trivially commutes with $S^{33}_{\pm}$.

Here, there is a subtle problem. 
The screening charges do not completely determine $\mathscr{F}_3(z)$ because they do not impose any constraint on the Heisenberg part. At least,  it is true in the CFT limit that the free field expression is the same for different values of $N$. We assume that is also true after q-deformation. One of the  reasonings for this assumption is that  the relevant quadratic relation in  the shifted toroidal algebra defined in \cite{Negut:2019agq} does not depend on the "shift". 
Then we have\footnote{We need to explain the notation. In (\ref{eq:F3}), we interpret $\rho(F_1)(z)$ as an operator acting on $\mathcal{F}^{(3)}\otimes\mathcal{F}^{(2)}\otimes\mathcal{F}^{(3)}\otimes\mathcal{F}^{(3)}$ by considering it acts on the last Fock space as an identity operator.}
\ba
\label{eq:F3}
\mathscr{F}_3(z)=\rho(F_1)(z).
\ea

Next, we consider the Heisenberg parts. We have two Heisenberg algebras, one of which is  the subalgebra of q-$W(\widehat{\mathfrak{gl}}_{3|1})$ while the other is q-$W(\widehat{\mathfrak{gl}}_1)$ itself.  As in the $U_q(\widehat{\mathfrak{gl}}_2)$ case, one of the linear combinations $J_n$ serves as  the element of the deformed $\mathcal{W}_3^{(2)}$ while the other one $j_n$  gives the $\mathfrak{gl}_1$  factor  which commute with all the elements of the deformed $\mathcal{W}_3^{(2)}$. We can determine their explicit forms from the commutation relation with $\mathscr{F}_3(z)$: the linear combination which commutes with $\mathcal{F}_3(z)$ is $j_n$.  To express them, we introduce the following boson,
\bq
\begin{split}
\frac{a_n^{\mathfrak{u}(1)}}{q^{-n}d^{-2n}-q^nd^{2n}}=
\begin{cases}
&-\frac{a_n^{(1)}}{d^n-d^{-n}}+\frac{d^na_n^{(2)}}{q_1^n-q_1^{-n}}-\frac{q^na_n^{(3)}}{d^n-d^{-n}}-\frac{q_3^{-n}a_n^{(4)}}{d^n-d^{-n}}\quad(n>0)\\
&-\frac{q_3^{-n}a_n^{(1)}}{d^n-d^{-n}}+\frac{d^{2n}a_n^{(2)}}{q_1^n-q_1^{-n}}-\frac{d^na_n^{(3)}}{d^n-d^{-n}}-\frac{a_n^{(4)}}{d^n-d^{-n}}\quad(n<0).
\end{cases}
\end{split}
\eq
We note that we have already used the same symbol in (\ref{eq:defanu1}). Here and below, we extend the definition of $a_n^{\mathfrak{u}(1)}$ to all $N$ by letting it be  the representation of $H_n$ as in  (\ref{eq:horizontalrep}). Then we have 

\ba
&&J_n=\begin{cases}
&\frac{(q_1^n-q_1^{-n})(q_3^n-q_3^{-n})d^{-n}}{(q^{-n}d^{-2n}-q^nd^{2n})^2}a_n^{\mathfrak{u}(1)}+\tilde{a}_n\quad(n>0)\\
&\frac{(q_1^n-q_1^{-n})(q_3^n-q_3^{-n})}{(q^{-n}d^{-2n}-q^nd^{2n})^2}a_n^{\mathfrak{u}(1)}+\tilde{a}_n\quad(n<0),
\end{cases}\\
\ \nonumber\\
&&j_n=\begin{cases}
&\frac{a_n^{\mathfrak{u}(1)}}{q^{-n}d^{-2n}-q^nd^{2n}}+\frac{\tilde{a}_n}{q_3^n-q_3^{-n}}\quad(n>0)\\
&\frac{a_n^{\mathfrak{u}(1)}}{q^{-n}d^{-2n}-q^nd^{2n}}+\frac{d^n\tilde{a}_n}{q_3^n-q_3^{-n}}\quad(n<0).
\end{cases}
\ea

Finally, we consider $\mathscr{E}_3(z)$. We will see that the following operator commutes with the screening charges:
\ba
\label{eq:E3}
\mathscr{E}_3(z)=\frac{1}{(q-q^{-1})^2}\left(e^{A(z)}-e^{B(z)}-e^{C(z)}+e^{D(z)}\right),
\ea
where
\small
\begin{equation}
\begin{split}
&A(z)=Q_{u_1}+u_{1,0}\log z-\frac{\epsilon_2}{2}u_{1,0}+\frac{1}{2}(3a_0^{(2)}-a_0^{(3)}-a_0^{(4)})-\sum_{n\neq0}\frac{A_n}{n}z^{-n},\\
&B(z)=Q_{u_1}+u_{1,0}\log z-\frac{\epsilon_2}{2}u_{1,0}+\frac{1}{2}(a_0^{(2)}+a_0^{(3)}-a_0^{(4)})-\sum_{n\neq0}\frac{B_n}{n}z^{-n},\\
&C(z)=Q_{u_1}+u_{1,0}\log z-\frac{\epsilon_2}{2}u_{1,0}+\frac{1}{2}(a_0^{(2)}-a_0^{(3)}+a_0^{(4)})-\sum_{n\neq0}\frac{C_n}{n}z^{-n},\\
&D(z)=Q_{u_1}+u_{1,0}\log z-\frac{\epsilon_2}{2}u_{1,0}-\frac{1}{2}(a_0^{(2)}-a_0^{(3)}-a_0^{(4)})-\sum_{n\neq0}\frac{D_n}{n}z^{-n},\\
\end{split}
\end{equation}
\normalsize
and
\begin{equation}
\begin{split}
&A_n=
\begin{cases}
-\frac{(q^n-q^{-n})d^{-3n}}{q_1^n-q_1^{-n}}a_n^{(2)}-\frac{(q^n-q^{-n})(q_1^n-q_1^{-n})d^{-3n}}{d^n-d^{-n}}a_n^{(3)}-\frac{(q^n-q^{-n})(q_1^n-q_1^{-n})d^{-2n}}{d^n-d^{-n}}a_n^{(4)}-\frac{q^n-q^{-n}}{q_3^n-q_3^{-n}}\tilde{a}_n\quad(n>0) \\
-\frac{(q^n-q^{-n})q^{-n}d^{-2n}}{q_1^n-q_1^{-n}}a_n^{(2)}-\frac{(q^n-q^{-n})q_3^n}{q_3^n-q_3^{-n}}\tilde{a}_n\qquad(n<0)
\end{cases}\\
&B_n=
\begin{cases}
-\frac{(q^n-q^{-n})d^{-n}}{q_1^n-q_1^{-n}}a_n^{(2)}-\frac{(q^n-q^{-n})(q_1^n-q_1^{-n})d^{-2n}}{d^n-d^{-n}}a_n^{(4)}-\frac{q^n-q^{-n}}{q_3^n-q_3^{-n}}\tilde{a}_n\qquad(n>0) \\
-\frac{(q^n-q^{-n})q^{-n}}{q_1^n-q_1^{-n}}a_n^{(2)}+\frac{(q^n-q^{-n})(q_1^n-q_1^{-n})d^{-2n}}{d^n-d^{-n}}a_n^{(3)}-\frac{(q^n-q^{-n})q_3^n}{q_3^n-q_3^{-n}}\tilde{a}_n\qquad(n<0)
\end{cases}\\
&C_n=
\begin{cases}
-\frac{(q^n-q^{-n})d^{-n}}{q_1^n-q_1^{-n}}a_n^{(2)}-\frac{(q^n-q^{-n})(q_1^n-q_1^{-n})d^{-n}}{d^n-d^{-n}}a_n^{(3)}-\frac{q^n-q^{-n}}{q_3^n-q_3^{-n}}\tilde{a}_n\qquad(n>0) \\
-\frac{(q^n-q^{-n})q^{-n}}{q_1^n-q_1^{-n}}a_n^{(2)}+\frac{(q^n-q^{-n})(q_1^n-q_1^{-n})d^{-n}}{d^n-d^{-n}}a_n^{(4)}-\frac{(q^n-q^{-n})q_3^n}{q_3^n-q_3^{-n}}\tilde{a}_n\qquad(n<0)
\end{cases}\\
&D_n=
\begin{cases}
-\frac{(q^n-q^{-n})d^n}{q_1^n-q_1^{-n}}a_n^{(2)}-\frac{q^n-q^{-n}}{q_3^n-q_3^{-n}}\tilde{a}_n\qquad(n>0) \\
-\frac{(q^n-q^{-n})q^{-n}d^{2n}}{q_1^n-q_1^{-n}}a_n^{(2)}+\frac{(q^n-q^{-n})(q_1^n-q_1^{-n})}{d^n-d^{-n}}a_n^{(3)}+\frac{(q^n-q^{-n})(q_1^n-q_1^{-n})d^{-n}}{d^n-d^{-n}}a_n^{(4)}-\frac{(q^n-q^{-n})q_3^n}{q_3^n-q_3^{-n}}\tilde{a}_n\qquad(n<0)
\end{cases}\\
\end{split}
\label{eq:BPboson}
\end{equation}
Here, the zero modes satisfy 
\ba
\label{eq:zeromodeidentity}
\epsilon_1u_{1,0}=\frac{1}{2}(a_0^{(1)}+a_0^{(3)}+a_0^{(4)}).
\ea
The definition of $a_n^{(4)}$ is the same as that of $a_n^{(1)}$ and $a_n^{(3)}$ given in (\ref{eq:subboson}).
We note that these vertex operators do not commute with $j_n$ for $n>0$. One can remove the $\mathfrak{gl}_1$ factor as follows:
\ba
\mathscr{E}_3(z)=\exp{\left(-\frac{1}{n}\sum_{n=1}^{\infty}\frac{q^nd^{2n}(q^n-q^{-n})(q_1^n-q_1^{-n})}{q_3^n+q_3^{-n}+q_1^{-n}}j_{-n}z^n\right)}\tilde{\mathscr{E}}_3(z),
\ea
where the new current $\tilde{\mathscr{E}}_3(z)$ does not contain $j_n$. However, the coefficients in (\ref{eq:BPboson}) becomes messy and that is why we include the extra factor. We note that it does not change the commutation relations because we add only the negative mode. 

One can see the commutativity with the screening charges for the first three Fock spaces by rewriting (\ref{eq:E3}) as follows:
\ba
e^{A(z)}-e^{B(z)}=\mathscr{E}_2(dz)\lambda_1(z),\quad e^{C(z)}-e^{D(z)}=\mathscr{E}_2(d^{-1}z)\lambda_2(z),
\ea
where we set
\ba
\mathscr{E}_2(z)=\rho(E_1(q_1z)).
\ea
Here, the vertex operators  $\lambda_1(z),\lambda_2(z)$ consist of the oscillators $a_n^{(4)}, \tilde{a}_n$ which trivially commute with the screening charges.
For the commutativity with $S^{33}_{\pm}$, we see that $e^{A(z)}$ and  $e^{D(z)}$ trivially commute with it because the boson oscillators in $A(z)$ and $D(z)$ are orthogonal to those in $S^{33}_{\pm}(z)$. The remaining term $e^{B(z)}+e^{C(z)}$ also commutes with $S^{33}_{\pm}$. To see that, we rewrite it as follows:
\small
\bq
\begin{split}
&e^{B(z)}+e^{C(z)}\\
=\ \ &\lambda_3(z)\Biggl(e^{\frac{1}{2}(a_0^{(3)}-a_0^{(4)})}\exp\Bigl(-\sum_{n<0}\frac{(q^n-q^{-n})(q_1^n-q_1^{-n})d^{-2n}}{n(d^n-d^{-n})}a_n^{(3)}z^{-n}\Bigr)\exp\Bigl(\sum_{n>0}\frac{(q^n-q^{-n})(q_1^n-q_1^{-n})d^{-2n}}{n(d^n-d^{-n})}a_n^{(4)}z^{-n}\Bigr)\Biggr.\\
&\Biggl.+e^{-\frac{1}{2}(a_0^{(3)}-a_0^{(4)})}\exp\Bigl(-\sum_{n<0}\frac{(q^n-q^{-n})(q_1^n-q_1^{-n})d^{-n}}{n(d^n-d^{-n})}a_n^{(4)}z^{-n}\Bigr)\exp\Bigl(\sum_{n>0}\frac{(q^n-q^{-n})(q_1^n-q_1^{-n})d^{-n}}{n(d^n-d^{-n})}a_n^{(3)}z^{-n}\Bigr)\Biggr)\\
=\ \ &e^{-\frac{1}{2}(a_0^{(3)}+a_0^{(4)})}\lambda_3(z)\rho(\Delta(E(z)))\exp\left(\sum_{n>0}\frac{(q^n-q^{-n})(q_1^n-q_1^{-n})d^{-n}}{n(d^n-d^{-n})}(a_n^{(3)}+d^na_n^{(4)})z^{-n}\right),
\end{split}
\eq
\normalsize
where  $\lambda_3(z)$ is the vertex operator which  trivially commutes with $S_{\pm}^{33}$. Here, we need to give some explanation for $\rho(\Delta(E(z)))$. The symbol $\Delta(E(z))$ means the coproduct for the current $E(z)$ of $\mathcal{E}_1$ defined in (\ref{eq:coprod}). The symbol $\rho$ denotes the representation on the third and fourth Fock spaces defined in  (\ref{eq:horizontalrep}). By construction, $\rho(\Delta(E(z)))$ commutes with $S_{\pm}^{33}$. The other factors trivially commute with $S_ {\pm}^{33}$.

As in the case of $\mathscr{F}_3(z)$, there is ambiguity in the Heisenberg part. We fixed it by demanding the Weyl symmetry  as follows. 
From the formulae in Appendix \ref{app:formulae}, we have
\small
\bq
\label{eq:E3E3}
\begin{split}
&(z-q^2w)\mathscr{E}_3(z)\mathscr{E}_3(w)+(w-q^2z)\mathscr{E}_3(w)\mathscr{E}_3(z)\\
=\ &q(1-q^2)(1-q_1^2)z\left(\delta\left(\frac{w}{d^2z}\right)(:e^{B(z)+C(w)}:-:e^{A(z)+D(w)}:)+d^{-2}\delta\left(\frac{d^2w}{z}\right)(:e^{C(z)+B(w)}:-:e^{D(z)+A(w)}:)\right)
\end{split}
\eq
\normalsize
Because of the relation
\ba
A_n+d^{-2n}D_n=B_n+d^{-2n}C_n,
\ea
the  RHS of (\ref{eq:E3E3}) vanishes and the relation (\ref{eq:E3E3}) is simplified as follows:
\ba
(z-q^2w)\mathscr{E}_3(z)\mathscr{E}_3(w)+(w-q^2z)\mathscr{E}_3(w)\mathscr{E}_3(z)=0.
\ea
This relation becomes the same as that of $\mathscr{F}_3(z)$ if we change $q$ to $q^{-1}$. That is interpreted as the Weyl transformation which flips the Dynkin diagram of $\mathfrak{gl}_{3|1}$.  

The quadratic relation between $\mathscr{E}_3(z)$ and $\mathscr{F}_3(w)$ is given as follows:
\bq
\label{eq:generatingeq}
\begin{split}
&[\mathscr{E}_3(z),\mathscr{F}_3(w)]\\
=\ &-\delta\left(\frac{w}{d^3z}\right)\frac{1-q_3^2}{(1-d^{-2})(q-q^{-1})^2}\mathscr{K}_3(w)+\delta\left(\frac{w}{dz}\right)\frac{1}{(q-q^{-1})^2}\mathscr{X}_3(w)-\delta\left(\frac{q^2dw}{z}\right)\frac{1}{(q-q^{-1})^2}\mathscr{\tilde{K}}^-_3(w),
\end{split}
\eq
where we define
\bq
\begin{split}
&\mathscr{K}_3(w)=:e^{A(d^{-3}w)+v_1^+(w)}:,\\
&\mathscr{X}_3(w)=q^{-2}:e^{A(d^{-1}w)+v_1^-(w)}:-\frac{1-q_1^2}{1-d^2}:e^{A(d^{-1}w)+v_1^+(w)}:+:e^{B(d^{-1}w)+v_1^+(w)}:+:e^{C(d^{-1}w)+v_1^+(w)}:,\\
&\mathscr{\tilde{K}}^-_3(w)=:e^{D(q^2dw)+v_1^-(w)}:.
\end{split}
\eq
The currents $\mathscr{K}_3(w)$ and $\tilde{\mathscr K}^-_3(w)$ are the vertex operators of the Heisenberg subalgebra as follows:
\ba
&&\mathscr{K}_3(w)=e^{k_0}:\exp{\left(-\sum_{n\neq0}\frac{1}{n}k_nw^{-n}\right)}:,\quad \mathscr{\tilde{K}}^-_3(w)=e^{\tilde{k}_0}\exp{\left(-\sum_{n\neq0}\frac{1}{n}\tilde{k}_nw^{-n}\right)},\\
&&k_n=d^{3n}A_n+v_{1,n}^+=\begin{cases}
&\frac{(q^n-q^{-n})(q^{-n}d^{-2n}-q^nd^{2n})}{q_3^n-q_3^{-n}}J_n\quad(n>0)\\
&-\frac{(q^n-q^{-n})(d^n-d^{-n}q_1^n)q_1^n}{q_3^n-q_3^{-n}}\tilde{a}_n\quad(n<0),
\end{cases}\\
&&\tilde{k}_n=q^{-2n}d^{-n}D_n+v_{1,n}^-=\begin{cases}
\qquad0\qquad\qquad(n>0)\\
-\frac{(q^n-q^{-n})(q_1^n-q_1^{-n})q_3^{2n}}{q^{-n}d^{-2n}-q^nd^{2n}}a_n^{u(1)}+\frac{(q^n-q^{-n})(q_1^n-q_1^{-n})d^{-n}}{q_3^n-q_3^{-n}}\tilde{a}_n\quad(n<0),
\end{cases}\\
&&k_0=-2\tilde{k}_0=-\frac{\epsilon_1-\epsilon_3}{\epsilon_1}(a_0^{(1)}+a_0^{(3)}+a_0^{(4)})+2a_0^{(2)}.
\ea
For later use, we split the current $\mathscr{K}_3(w)$ into the positive and negative parts as follows:
\ba
&&\mathscr{K}_3(w)=\mathscr{K}^-_3(w)\mathscr{K}^+_3(w),\\
&&\mathscr{K}^-_3(w)=e^{k_0}:\exp{\left(\sum_{n>0}\frac{1}{n}k_{-n}w^n\right)},\qquad\mathscr{K}^+_3(w)=\exp{\left(-\sum_{n>0}\frac{1}{n}k_nw^{-n}\right)}.
\ea
The current $\mathscr{X}_3(w)$ can be written as follows:
\bq
\begin{split}
\mathscr{X}_3(w)=&\exp{\left(\sum_{n=1}^{\infty}\frac{q_3^{-n}(q^n-q^{-n})(q_1^n-q_1^{-n})(d^n-d^{-n})}{n(q^{-n}d^{-2n}-q^nd^{2n})^2}a_{-n}^{u(1)}w^n\right)}\rho\bigl(\Delta^{(4)}(t(d^{-1}w))\bigr)\\
&\hspace{200pt}e^{\tilde{a}_0}\exp{\left(-\sum_{n=1}^{\infty}\frac{(q^n-q^{-n})q^{-n}}{n}J_nw^{-n}\right)},
\end{split}
\eq
where the current $t(z)$ is defined in (\ref{eq:tnewcurrent}). Here, we introduce the notation for the higher coproduct $\Delta^{(n+1)}=(1\otimes\Delta)\Delta^{(n)}$, $\Delta^{(2)}=\Delta$. The symbol $\rho$ denotes the representation on $\mathcal{F}^{(3)}\otimes\mathcal{F}^{(2)}\otimes\mathcal{F}^{(3)}\otimes\mathcal{F}^{(3)}$. 
We can interpret it as q-analogue of Virasoro generator. 

It is convenient to change the normalization as follows to make the the Weyl symmetry clear,
\bq
\begin{split}
&G^+(z)=\mathscr{E}_3(d^{-3}z),\quad G^-(z)=\mathscr{F}_3(z),\quad\tilde{\psi}^-(z)=\tilde{\mathscr{K}}^-_3(q^{-1}d^{-2}z),\quad T(z)=\mathscr{X}_3(d^{-1}z)\\
&\psi(z)=\psi^-(z)\psi^+(z),\quad\psi^-(z)=q^{-1}\mathscr{K}^-_3(z),\quad\psi^+(z)=\mathscr{K}^+_3(z).
\end{split}
\eq
Then the equation (\ref{eq:generatingeq}) is rewritten into
\bq
\label{eq:GGrelation}
\begin{split}
&[G^+(z),G^-(w)]\\
=\ &-\delta\left(\frac{w}{z}\right)\frac{q(1-q_3^2)}{(1-d^{-2})(q-q^{-1})^2}\psi(w)+\delta\left(\frac{d^2w}{z}\right)\frac{1}{(q-q^{-1})^2}T(dw)-\delta\left(\frac{q^2d^4w}{z}\right)\frac{1}{(q-q^{-1})^2}\tilde{\psi}^-(qd^2w).
\end{split}
\eq
The quadratic relations  between two of these currents are given as follows:
\ba
&&\psi(z)G^-(w)-q^2\frac{1-\frac{z}{q^2w}}{1-\frac{q^2z}{w}}G^-(w)\psi(z)\label{eq:psiGm}\\
&=&-\frac{(1-q^{-4})(1-q^4d^2)}{1-q_3^{-2}}\delta\left(\frac{w}{q^2z}\right)\psi^-(z)G^-(w)\psi^+(z)+\frac{(1-q^2)(1-q_1^2)}{1-q_3^{-2}}\delta\left(\frac{d^2w}{z}\right)\psi^-(z)G^-(w)\psi^+(z),\nonumber\\
&& \nonumber\\
&&\tilde{\psi}^-(z)G^-(w)=q^2\frac{1-\frac{z}{q^3d^2w}}{1-\frac{z}{q^{-1}d^2w}}G^-(w)\tilde{\psi}^-(z),\label{eq:psitGm}\\
&& \nonumber\\
&&q^2\frac{1-\frac{w}{q^2d^{-1}z}}{1-\frac{q^2dw}{z}}T(z)G^-(w)-G^-(w)T(z)=\frac{(1-q_1^2)(q-q^{-1})}{1-d^2}\delta\left(\frac{w}{dz}\right)\psi^-(w)G^-(d^{-2}w)\psi^+(w),\label{eq:TGm}\\
&& \nonumber\\
&&\psi(z)G^+(w)-q^{-2}\frac{1-\frac{q^2z}{w}}{1-\frac{z}{q^2w}}G^+(w)\psi(z)\label{eq:psiGp}\\
&=&-\frac{(1-q^4)(1-q^{-4}d^{-2})}{1-q_3^2}\delta\left(\frac{q^2w}{z}\right)\psi^-(z)G^+(w)\psi^+(z)+\frac{(1-q^{-2})(1-q_1^{-2})}{1-q_3^2}\delta\left(\frac{w}{d^2z}\right)\psi^-(z)G^+(w)\psi^+(z),\nonumber\\
&& \nonumber\\
&&\tilde{\psi}^-(z)G^+(w)=q^{-2}\frac{1-\frac{q^3d^2z}{w}}{1-\frac{q^{-1}d^2z}{w}}G^+(w)\tilde{\psi}^-(z),\label{eq:psitGp}\\
&& \nonumber\\
&&q^{-2}\frac{1-\frac{q^2d^{-1}w}{z}}{1-\frac{q^{-2}d^{-1}w}{z}}T(z)G^+(w)-G^+(w)T(z)=-\frac{(1-q_1^{-2})(q-q^{-1})}{1-d^{-2}}\delta\left(\frac{dw}{z}\right)\psi^-(w)G^+(d^2w)\psi^+(w),\label{eq:TGp}
\ea
One can see again that these relations are invariant under the permutation of the currents, $G^+(z)$ and $G^-(z)$, and the inversion of the parameters $q_i\to q_i^{-1}$ $(i=1,2,3)$.

\subsection{General $N$}
We consider the cases for general $N$. We consider  $\mathcal{F}^{(3)}\otimes\mathcal{F}^{(2)}\otimes(\mathcal{F}^{(3)})^{\otimes N-1}$ for the Fock space of q-$W(\widehat{\mathfrak{gl}}_{N|1})$. As in the previous case, the current $\mathscr{F}_N(z)$ is given by
\ba
\mathscr{F}_N(z)=\rho(F_1(z)).
\ea
For the current $\mathscr{E}_N(z)$, we claim that it is possible to decouple from $\mathscr{E}_N(z)$ the vertex operator $\Lambda(z)$ acting on $\tilde{\mathcal{F}}^{(3)}$ as follows:

\ba
&&\mathscr{E}_N(z)=\mathscr{E'}_N(z)\Lambda(d^{3-N}z),\\
&&\Lambda(z)=\exp\left(-\sum_{n>0}\frac{(q^n-q^{-n})q_3^{-n}}{n(q_3^n-q_3^{-n})}\tilde{a}_{-n}z^n\right)\exp\left(\sum_{n>0}\frac{q^n-q^{-n}}{n(q_3^n-q_3^{-n})}\tilde{a}_nz^{-n}\right).
\ea
The vertex operator $\mathscr{E'}_N(z)$ is determined by the following recursion relation:
\ba
&&\mathscr{E'}_{N+1}(z)=\mathscr{E'}_N(dz)e^{\phi^+_{(N+2)}(z)}-e^{\phi^-_{(N+2)}(z)}\mathscr{E'}_N(d^{-1}z)\label{eq:recursionE},
\ea
where 
\ba
&&\phi^+_{(m)}(z)=-\frac{1}{2}a_0^{(m)}+\tilde{a}_0+\sum_{n>0}\frac{(q^n-q^{-n})(q_1^n-q_1^{-n})d^{-2n}}{n(d^n-d^{-n})}a_n^{(m)}z^{-n}\hspace{30pt}\label{eq:psimplus}\\
&&\phi^-_{(m)}(z)=-\sum_{n>0}\frac{(q^n-q^{-n})(q_1^n-q_1^{-n})d^n}{n(d^n-d^{-n})}a_{-n}^{(m)}z^n+\frac{1}{2}a_0^{(m)} \label{eq:psimminus}
\ea
for $m\geq3$. We note that we have changed the normalization of $\mathscr{E}_N(z)$ for simplicity. The zero modes  satisfy the condition, 
\ba
\epsilon_1u_{1,0}=\frac{1}{2}\sum_{\substack{j=1\\j\neq2}}^{N+1}a_0^{(j)}.
\ea
The initial condition is given by 
\ba
&&\mathscr{E'}_1(z)=:e^{\phi_{(2)}(z)}:,\\
&&\phi_{(2)}(q_3^{1/2}z)=Q_{u_1}+u_{1,0}\log z+\frac{3\epsilon_1}{2}u_{1,0}-\frac{1}{2}a_0^{(2)}+\sum_{n\neq0}\frac{(q^n-q^{-n})q_1^{-\frac{|n|}{2}}}{n(q_1^n-q_1^{-n})}a_n^{(2)}z^{-n}.
\ea
We note that the $N=1$ case  corresponds to q-analogue of $\beta\gamma$ CFT. 

By induction, we can show that $\mathscr{E}_{N+1}(z)$ commutes with these screening charges in the following way. We assume the claim is true for $\mathscr{E}_N(z)$.
From  the recursion relation (\ref{eq:recursionE}), it is obvious that the current $\mathscr{E}_{N+1}(z)$ commutes with the screening charges except the last one $S_{\pm}^{33}$.  To show the commutativity with the remaining one, we use the explicit form of the current,
\ba
&&\mathscr{E'}_N(z)=\sum_{s_3=\pm1}\cdots\sum_{s_{N+1}=\pm1}(-1)^{P_s}:e^{\phi^{s_3,s_4\cdots,s_{N+1}}(z)}:,\\
&&\phi^{s_3,s_4\cdots,s_{N+1}}(z)=\phi_{(2)}(d^{\sum_{i=3}^{N+1}s_i}z)+\sum_{j=3}^{N+1}\phi_{(j)}^{s_j}(d^{\sum_{k=j+1}^{N+1}s_k}z),
\ea
where we use the notation,
\ba
P_s=&\sum_{i=3}^{N+1}\frac{1-s_i}{2},\qquad
\phi_{(i)}^p=\begin{cases}
\phi_{(i)}^+\qquad p=1\\
\phi_{(i)}^-\qquad p=-1
\end{cases}
\quad(i\geq3).
\ea
By rewriting this into
\bq
\begin{split}
\mathscr{E'}_N(z)=&\sum_{s_3=\pm1}\cdots\sum_{s_{N-1}=\pm1}(-1)^{P_s}\left(:e^{\phi^{s_3,s_4\cdots,+,+}(z)}:+:e^{\phi^{s_3,s_4\cdots,-,-}(z)}:\right.\\
&\left.\hspace{70pt}-:e^{\phi^{s_3,s_4\cdots,s_{N-1}}(z)}:(:e^{\phi^-_{(N)}(dz)+\phi^+_{(N+1)}(z)}:+:e^{\phi^+_{(N)}(d^{-1}z)+\phi^-_{(N+1)}(z)}:)\right),
\end{split}
\eq
we can see the commutativity in the same manner as in the $N=3$ case. We note that  the zero modes $\tilde{a}_0=a_0^{(2)}+\frac{\epsilon_3-\epsilon_1}{2\epsilon_1}\sum_{\substack{ j=1\\ j\neq 2}}^{m}a_0^{(j)}$  included in (\ref{eq:psimplus}) do not play any role in the relation to the screening charges, but they are necessary to produce the desired quadratic relations as we will see in the following.

Let us derive the quadratic relations. We first consider the one between $\mathscr{E}_N(z)$ and $\mathscr{E}_N(w)$. One can show by induction that it is given by 
\ba
\label{eq:ENEN}
(z-q_2w)\mathscr{E}_N(z)\mathscr{E}_N(w)=(w-q_2z)\mathscr{E}_N(w)\mathscr{E}_N(z).
\ea
For the detail of the computation, see Appendix \ref{app:proof}.

Next, we consider the relation between $\mathscr{E}_N(z)$ and $\mathscr{F}_N(z)$. We set 
\ba
\mathscr{E}_N^{s_3,s_4\cdots s_{N+1}}(z)=:e^{\phi^{s_3,s_4\cdots,s_{N+1}}(z)}:\Lambda(d^{3-N}z).
\ea
Then we have
\bq
\begin{split}
[\mathscr{E}_N(z),\mathscr{F}_N(w)]=&(-1)^N\delta\left(\frac{q^2d^{N-2}w}{z}\right)\tilde{\mathscr K}_N^-(w)-q^{4-2N}\delta\left(\frac{w}{d^Nz}\right)\Biggl(\prod_{\substack{j=0\\ j\neq i}}^{N-2}\frac{1-q^2d^{2(i-j)}}{1-d^{2(i-j)}}\Biggr)\mathscr{K}_N(w)\\
&\hspace{120pt}-\displaystyle{\sum_{i=0}^{N-3}\delta\left(\frac{d^{N-4-2i}w}{z}\right)}\mathscr{X}_N^{(N-1-i)}(w),
\end{split}
\label{eq:ENFNquad}
\eq
where
\ba
&&\tilde{\mathscr K}_N^-(w)=:\mathscr{E}_N^{s_3=s_4=\cdots=s_{N+1}=-1}(q^2d^{N-2}w)e^{v_1^-(w)}:\\
&&\mathscr{K}_N(w)=:\mathscr{E}_N^{s_3=s_4=\cdots=s_{N+1}=1}(d^{-N}w)e^{v_1^+(w)}:\\
&&\mathscr{X}_N^{(N-1-i)}(w)=\hspace{-4mm}\displaystyle{\sum_{\substack{s_3,\cdots ,s_{N+1}\\P_s\leq N-2-i}}}\hspace{-3mm}(-1)^{P_s}q^{2P_s+4-2N}\Biggl(\prod_{\substack{j=0\\ j\neq i}}^{N-2-P_s}\frac{1-q^2d^{2(i-j)}}{1-d^{2(i-j)}}\Biggr):\mathscr{E}_N^{s_3,s_4\cdots s_{N+1}}(d^{N-4-2i}w)e^{v_1^+(w)}: \nonumber\\
&&\hspace{50pt}-\hspace{-5mm}\displaystyle{\sum_{\substack{s_3,\cdots,s_{N+1}\\P_s\leq N-3-i}}\hspace{-3mm}(-1)^{P_s}q^{2P_s+4-2N}\Biggl(\prod_{\substack{j=0\\ j\neq i}}^{N-3-P_s}\frac{1-q^2d^{2(i-j)}}{1-d^{2(i-j)}}\Biggr)}:\mathscr{E}_N^{s_3,s_4\cdots s_{N+1}}(d^{N-4-2i}w)e^{v_1^-(w)}:.
\ea
This relation can be derived from  (\ref{eq:Ncommutator1})  and  (\ref{eq:Ncommutator2}).
The currents 
\ba
\mathscr{K}_N(w)=e^{k_0}\exp\left(-\sum_{n\neq0}\frac{1}{n}k_nw^{-n}\right),\qquad\tilde{\mathscr K}_N^-(w)=e^{\tilde{k}_0}\exp\left(-\sum_{n\neq0}\frac{1}{n}\tilde{k}_nw^{-n}\right)
\ea
 are the vertex operators for the Heisenberg subalgebras as before. To see it, we introduce the $\mathfrak{u}(1)$ boson,
\bq
\begin{split}
\frac{a_n^{\mathfrak{u}(1)}}{q^{-n}d^{-(N-1)n}-q^nd^{(N-1)n}}=
\begin{cases}
-\frac{a_n^{(1)}}{d^n-d^{-n}}+\frac{d^na_n^{(2)}}{q_1^n-q_1^{-n}}-\sum_{m=3}^{N+1}\frac{q^nd^{(m-3)n}a_n^{(m)}}{d^n-d^{-n}}\quad(n>0)\\
-\frac{q^nd^{(N-2)n}a_n^{(1)}}{d^n-d^{-n}}+\frac{d^{(N-1)n}a_n^{(2)}}{q_1^n-q_1^{-n}}-\sum_{m=3}^{N+1}\frac{d^{(N+1-m)n}a_n^{(m)}}{d^n-d^{-n}}\quad(n<0).
\end{cases}
\end{split}
\eq
Then we have
\ba
&&k_n=\begin{cases}
(1-q^{-2n})\left(\frac{(q_1^n-q_1^{-n})}{q^{-n}d^{-(N-1)n}-q^nd^{(n-1)n}}a_n^{\mathfrak{u}(1)}+\frac{q_3^n-q_3^{-n}d^{(2N-4)n}}{q_3^n-q_3^{-n}}\tilde{a}_n\right)\qquad(n>0)\\
\frac{(1-q^{-2n})(1-d^{(2N-4)n})}{q_3^n-q_3^{-n}}\tilde{a}_n\qquad(n<0),
\end{cases}\\
&&\tilde{k}_n=\begin{cases}
\qquad 0\qquad\quad(n>0)\\
-(q^n-q^{-n})q^{-2n}\left(\frac{(q_1^n-q_1^{-n})d^{(1-N)n}}{q^{-n}d^{-(N-1)n}-q^nd^{(N-1)n}}a_n^{\mathfrak{u}(1)}+d^{-n}\tilde{a}_n\right)\qquad(n<0),
\end{cases}\\
&&\tilde{k}_0=-\frac{1}{N-1}k_0=-a_0^{(2)}+\frac{\epsilon_1-\epsilon_3}{2\epsilon_1}\sum_{\substack{i=1\\ i\neq2}}^{N+1}a_i.
\ea
Apart from the Heisenberg part, we have $N-2$ currents in the RHS of (\ref{eq:ENFNquad}). 
That is the same as the undeformed case \cite{Feigin:2004wb}. 
The currents $\mathscr{X}_N^{(n)}$ can be seen as q-analogue of the $W_n$ current.

Finally, we mention the representation associated with another Dynkin diagram of $\mathfrak{gl}_{N|1}$. So far, we have fixed the order of the Fock space to 
$\mathcal{F}^{(3)}\otimes\mathcal{F}^{(2)}\otimes(\mathcal{F}^{(3)})^{\otimes N-1}$, but it can be easily  generalized to another ordering. As one can see from (\ref{eq:upm}), the current $\mathscr{E}_N(z)$ does not act on the first Fock space, which implies that it commutes with the screening charges of  $(\mathcal{F}^{(3)})^{\otimes M}\otimes\mathcal{F}^{(2)}\otimes(\mathcal{F}^{(3)})^{\otimes N-1}$ without any modification. For the current $\mathscr{F}_N(z)$, we need modification, but that can be similarly done  as previously discussed.  We expect that the quadratic relations do not depend on the ordering of the Fock space.

\section{The norm of  Whittaker states and Nekrasov's instanton partition function}
\label{sec:5dAGT}
In  the AGT correspondence, non-principal W-algebras appear as the dual of 4d $\mathcal{N}=2$ supersymmetric gauge theories with surface operators.
The half-BPS surface operators in SU($N$) gauge theories are labelled by the partition of $N$, which represents the preserved gauge symmetry. For the partition $[n_1,n_2\cdots n_M]$, the preserved  gauge group is 
\ba
S[\U(n_1)\times \U(n_2)\times\cdots\times \U(n_M)].
\label{eq:preserved_group}
\ea 
In 2d side, the partition is related with $\mathfrak{su}(2)$ embedding in the Drinfeld-Sokolov reduction of $\widehat{\mathfrak{su}}(N)$  \cite{Wyllard:2010rp,Wyllard:2010vi,Kanno:2011fwx}. 
Feigin-Semikhatov's W-algebra is associated with the partition $N=(N-1)+1$ and the corresponding surface operator is called simple. 

It is considered that the above correspondence lifts to 5d $\mathcal{N}=1$ supersymmetric gauge theory on $\mathbb{C}_{\epsilon_1}\times\mathbb{C}_{\epsilon_2}\times S^1$ and the corresponding chiral algebra should lift to its deformation. In this section, we focus on q-deformed Bershadsky-Polyakov algebra and examine whether it matches with 5d AGT by comparing the norm of Whittaker states (Gaiotto states)  with the instanton partition function. 

\subsection{Nekrasov's partition function under the existence of a surface operator}
In this section, we  review  the instanton partition function with a surface operator following \cite{Kanno:2011fwx}.  In the following, we consider $\U(N)$ instead of $\SU(N)$. 
Let us first see the case without a surface operator. The moduli space is given by the ADHM data. To describe it, we introduce the $k$-dimensional vector space $V$ and $N$-dimensional vector space $W$  for $k$-instantons in $\U(N)$ gauge theory. Then, the moduli space is described by the data $B_1,B_2\in{\rm Hom}(V,V),I\in {\rm Hom}(W,V),J\in{\rm Hom}(V,W)$ with ADHM constraint imposed on. The theory has the rotational symmetry SO(4) and the gauge symmetry $\U(N)$.  The action of their Cartan torus $\U(1)^{N+2}$ on the moduli space is the key ingredient for the computation of the instanton partition function.  Concretely, the contribution to the instanton partition function comes only from the fixed points of the torus action labelled by $N$ Young diagrams  and  can be evaluated from the character of the tangent spaces there.  The character of the vector spaces $V,W$ at the fixed point $\vec{Y}=(Y_1,\cdots Y_N)$ are given as follows:
\ba
W=\sum_{\alpha=1}^Nu_\alpha,\quad V=\sum_{\alpha=1}^Nu_{\alpha}\sum_{(i,j)\in Y_{\alpha}}q_1^{1-i}q_2^{1-j},
\label{eq:characterWV}
\ea
where $u_{\alpha}=e^{a_\alpha}$ and $q_{1,2}=e^{\epsilon_{1,2}}$ are the equivalent parameters\footnote{In this subsection, we use the parameters $q_{1,2}$ independently of the quantum toroidal algebra.}.
Here, we abuse the notation and identify the vector spaces with their characters. 
The character of the tangent space is given by 
\ba
\chi_{\vec{Y}}=-(1-q_1)(1-q_2)V^*V+W^*V+q_1q_2V^*W,
\label{eq:character_tang}
\ea
where the parameters $u_\alpha, q_i$ are replaced with their inverse in $V^*$ and $W^*$.  Substituting (\ref{eq:characterWV}) into (\ref{eq:character_tang}),  one finds that the character can be expressed as  
\ba
\chi_{\vec{Y}}=\sum_{i=1}^{2Nk}e^{w_i(\vec Y)}.
\ea
Then the Nekrasov partition function \cite{Nekrasov2004} for 5d SYM is given by
\ba
Z=\sum_{\vec Y}x^{|\vec{Y}|}\frac{1}{\prod_{i=1}^{2Nk}(1-e^{w_i(\vec Y)})}.
\ea
The 4d case is obtained  by rescaling the parameters $a_\alpha\to Ra_\alpha$, $\epsilon_{1,2}\to R\epsilon_{1,2}$ and taking the limit $R\to0$.

Next, we  review  the instanton partition function with a surface operator. 
The crucial point in \cite{Kanno:2011fwx} is that the instanton moduli space under the existence of a surface operator labelled by $[n_1,n_2,\cdots n_M]$ can be described as that on the orbifold $\mathbb{C}_{\epsilon_1}\times\mathbb{C}_{\epsilon_2}/\mathbb{Z}_M$. Due to the orbifolding, it is natural to rescale $q_2$ to $q_2^{1/M}$ and redefine $u_{1,\cdots N}$ as $u_{s,I}$, where $s=1,\cdots n_I$ and $I=1,\cdots  M$. The vector spaces $V,W$ are decomposed into the subspaces by the $\mathbb{Z}_M$ charges,
\ba
W=\oplus_{I=1}^MW_I, \quad V= \oplus_{I=1}^MV_I.
\ea
The vector space $W_I$ is $n_I$-dimensional and  its character is given as follows:
\ba
W_I=\sum_{s=1}^{n_I}q_2^{-I/M}u_{s,I}.
\ea
The vector space $V_I$ consists of the vector corresponding to the box $(i,j')\in Y_{s,I-j'+1}$ and its character is given by
\ba
V_I=\sum_{J=1}^M\sum_{s=1}^{n_{I-J+1}}q_2^{\lfloor\frac{I-J}{M}\rfloor-I/M}u_{s,I-J+1}\sum_{(i,jM+J)\in Y_{s,I-J+1}}q_1^{1-i}q_2^{-j},
\ea
where $\lfloor\frac{I-J}{M}\rfloor=\begin{cases}\ \ 0\quad(I\geq J)\\-1\quad(I<J)\end{cases}$.
We note that the index  should be understood modulo $M$.
The fixed point is still labelled by $N$ Young diagrams, but the dimension of the tangent space becomes smaller because the ADHM data is restricted to 
\ba
B_1\in{\rm Hom}(V_I,V_I),\quad B_2\in{\rm Hom}(V_I,V_{I+1}),\quad I\in {\rm Hom}(W_I,V_I),\quad J\in{\rm Hom}(V_I,W_{I+1}).
\ea
Then the character of the tangent space is given by
\ba
\chi_{\vec Y}=\sum_{I=1}^M\left(-(1-q_1)V_I^*V_I+(1-q_1)q_2^{1/M}V_{I-1}^*V_I+W_I^*V_I+q_1q_2^{1/M}V_{I-1}^*W_I\right).
\ea
The instanton partition function is obtained in the similar way with the previous case, but we should change the fugacity parameter $x$ to $\prod_{I=1}^Mx_I^{d_I(\vec Y)}$, where $d_I(\vec Y)$ denotes the dimension of $V_I$.  These fugacity parameters measure  the magnetic charges associated with the $\U(1)$ subgroups in (\ref{eq:preserved_group}) as well as the instanton number.

As an example, we consider the $\U(N)$ instanton partition function under the presence of the surface operator of the type $[N-1,1]$. We focus on the coefficients of $x_1^{d_1}$ and $x_2^{d_2}$. For $x_2^{d_2}$, the Young diagrams except the last one must be empty because of $d_1=0$. The last one consists of only one column.
For such Young diagram, the character of each vector space is  given as follows:
\ba
W_1=\sum_{s=1}^{N-1}u_{s,1}q_2^{-1/2},\quad W_2=u_{1,2}q_2^{-1},\quad V_1=0,\quad V_2=\sum_{i=1}^{d_2}q_1^{1-i}q_2^{-1}u_{1,2},
\ea
which leads to
\ba
\chi_{\vec Y}=\sum_{i=1}^{d_2}q_1^i\left(1+\sum_{s=1}^{N-1}\frac{q_2u_{s,1}}{u_{1,2}}\right).
\ea
Then the instanton partition function is given by
\bq
Z^{(S)}(\vec{u},(\emptyset,\cdots\emptyset,1^{d_2}))
=\Biggl(\prod_{i=1}^{d_2}\Bigl((1-q_1^i)\prod_{s=1}^{N-1}(1-\frac{u_{s,1}}{u_{1,2}}q_1^iq_2)\Bigr)\Biggr)^{-1}.
\label{eq:Zsurface1}
\eq

For $x_1^{d_1}$, the computation becomes more complicated. In this case, the last Young diagram is empty while all of the others are in the shape of one-column Young diagrams. For simplicity, we consider the case of $N=3$. After some computations,  one can see that the final expression can be written in a simple form. From the result for $d_1=1,2,3,4$, we conjecture the following form:
\ba
\sum_{n=0}^{d_1}Z^{(S)}(\vec{u},(1^n,1^{d_1-n},\emptyset))=\Biggl(\prod_{i=1}^{d_1}\Bigl(1-q_1^i\Bigr)\Bigl(1-\frac{u_{1,2}}{u_{1,1}}q_1^i\Bigr)\Bigl(1-\frac{u_{1,2}}{u_{2,1}}q_1^i\Bigr)\Biggr)^{-1}.
\label{eq:Zsurface2}
\ea
The similarity between (\ref{eq:Zsurface1}) and (\ref{eq:Zsurface2}) can be understood in terms of the chiral algebra. It corresponds to the Weyl transformation which we discussed in the previous section.

\subsection{Comparison}
In this subsection, we compare the norm of the Whittaker states for the deformed Bershadsky-Polyakov algebra with the instanton partition function of  5d $\mathcal{N}=1$ U(3) SYM  with the surface operator of the type $3=2+1$.  We can define the Whittaker state in the same way as the 4d case \cite{Wyllard:2010rp,Kanno:2011fwx}. 

We set the mode expansion as follows:
\bq
\begin{split}
&G^+(z)=e^{u_{1,0}\log z}\sum_{n\in\mathbb{Z}}G_n^+z^{-n},\quad G^-(z)=e^{-u_{1,0}\log z}\sum_{n\in\mathbb{Z}}G_n^-z^{-n},\quad T(z)=\sum_{n\in\mathbb{Z}}T_nz^{-n},\\&\psi(z)=\sum_{n\in\mathbb{Z}}\psi_nz^{-n},\quad  \tilde{\psi}(z)=\sum_{n\geq0}\tilde{\psi}_{-n}z^n.
\end{split}
\eq
The highest weight state is defined by the conditions,
\bq
\begin{split}
&G_{n-1}^+\ket{\rm hw}=G_n^-\ket{\rm hw}=T_n\ket{\rm hw}=k_n\ket{\rm hw}=0,\quad{\rm for\ }n\geq1,\\
&\bra{\rm hw}G_{-n+1}^-= \bra{\rm hw}G_{-n}^+= \bra{\rm hw}T_{-n}= \bra{\rm hw}k_{-n}=0,\quad{\rm for\ } n\geq1,\\
&T_0\ket{\rm hw}=h\ket{\rm hw},\quad e^{k_0}\ket{\rm hw}=J\ket{\rm hw}.
\end{split}
\eq
The Whittaker state  is then defined by the conditions,
\bq
\begin{split}
&G^+_n\ket{W}=t_1^{1/2}\delta_{n,0}\ket{W},\quad \bra{W}G^-_{-n}=\bra{W}t_1^{1/2}\delta_{n,0},\quad{\rm for\ } n\geq0,\\
&G^-_n\ket{W}=t_2^{1/2}\delta_{n,1}\ket{W},\quad \bra{W}G^+_{-n}=\bra{W}t_2^{1/2}\delta_{n,1},\quad{\rm for\ } n\geq1,\\
&T_n\ket{W}=k_n\ket{W}=0,\qquad \bra{W}T_{-n}=\bra{W}k_{-n}=0\quad{\rm  for\ } n\geq1,
\end{split}
\eq
where it is assumed that the Whittaker state can be expanded by the descendant states  of a single highest weight $\ket{\alpha,\beta}$ state as follows:
\ba
&&\ket{W}=\sum_{n,m\geq0}t_1^{n/2}t_2^{m/2}\ket{n,m,\alpha,\beta},\\
&&\bra{W}=\sum_{n,m\geq0}t_1^{n/2}t_2^{m/2}\bra{n,m,\alpha,\beta}.
\ea
Here, the normalization is fixed by
\ba
&&\ket{0,0,\alpha,\beta}=\ket{\alpha,\beta},\\
&&\braket{\alpha,\beta|\alpha,\beta}=1,
\ea
and the parameters $\alpha,\beta$ are associated with $h,J$ as we will see below. Each term in the above expansion is expected to be fixed recursively. In general, it is a tedious work to compute the states at higher order, but we can easily obtain the states for $n=0$ or $m=0$ because the states with such charges are uniquely fixed to
\ba
&&\ket{n,0,\alpha,\beta}\propto(G_{0}^-)^n\ket{\alpha,\beta},\qquad \bra{n,0,\alpha,\beta}\propto\bra{\alpha,\beta}(G_0^+)^n\\
&&\ket{0,m,\alpha,\beta}\propto(G_{-1}^+)^m\ket{\alpha,\beta},\qquad \bra{n,0,\alpha,\beta}\propto\bra{\alpha,\beta}(G_1^-)^n.
\ea
The proportional factors are immediately read off from the conditions such as $(G_0^+)^n\ket{n,0,\alpha,\beta}=\ket{\alpha,\beta}$. Then the norm of the Whittaker states is written as 
\ba
\label{eq:Wnorm}
\braket{W|W}=1+\sum_{n=1}^{\infty}\frac{t_1^n}{\braket{\alpha,\beta |(G_0^+ )^n(G_0^-)^n|\alpha,\beta}}+\sum_{m=1}^{\infty}\frac{t_2^m}{\braket{\alpha,\beta |(G_1^- )^m(G_{-1}^+)^m|\alpha,\beta}}+\cdots.
\ea
In the following, we compare (\ref{eq:Wnorm})  with the instanton partition function (\ref{eq:Zsurface1}) and (\ref{eq:Zsurface2}). 

We first compute the coefficient of $t_1^n$. The easiest way is to use the free field representation. For that, we need to describe the highest weight state in terms of the free bosons. The free boson module is defined as follows:
\ba
&&a_n^{(i)}\ket{\alpha,\beta,\gamma,\delta}=0,\quad \tilde{a}_{n}\ket{\alpha,\beta,\gamma,\delta}=0,\quad(n>0,\ i=1,2,3,4)\\
&&\frac{1}{2}(3a_0^{(2)}-a_0^{(3)}-a_0^{(4)})\ket{\alpha,\beta,\gamma,\delta}=\alpha\ket{\alpha,\beta,\gamma,\delta},\\
&&\frac{1}{2}(a_0^{(2)}+a_0^{(3)}-a_0^{(4)})\ket{\alpha,\beta,\gamma,\delta}=\gamma\ket{\alpha,\beta,\gamma,\delta},\\
&&\frac{1}{2}(a_0^{(2)}-a_0^{(3)}+a_0^{(4)})\ket{\alpha,\beta,\gamma,\delta}=\delta\ket{\alpha,\beta,\gamma,\delta},\\
&&\frac{1}{2}(a_0^{(2)}-a_0^{(1)})\ket{\alpha,\beta,\gamma,\delta}=\beta\ket{\alpha,\beta,\gamma,\delta}.
\ea
We also impose
\ba
\label{eq:picturefix}
u_{1,0}\ket{\alpha,\beta,\gamma,\delta}=0.
\ea
We note that the condition $u_{1,0}=0$ holds in the whole Fock space because of $[Q_{u_1},u_{1,0}]=0$. 
This process is so-called picture fixing. Combining it with (\ref{eq:zeromodeidentity}), we have
\ba
a_0^{(1)}+a_0^{(3)}+a_0^{(4)}=0,
\ea
which is rewritten into
\ba
\alpha+\beta=2(\gamma+\delta).
\ea
The state  $\ket{\alpha,\beta,\gamma,\delta}$ satisfies all the anihilation conditions except the one for $G_0^+$. 
It requires the relation
\ba
(e^{\alpha}-e^{\gamma})(1-e^{\delta-\alpha})=0.
\ea
We have the two solutions, $\alpha=\gamma$ and $\alpha=\delta$, but the difference does not cause any effect on the representation theory. From the above constraint, all the parameters can be expressed by $\alpha$ and $\beta$. In the following, we denote the highest weight state just by $\ket{\alpha,\beta}$. The eigenvalues $h,J$ are given by 
\ba
h=q^{-2}e^{\alpha-\beta}-\frac{1-q_1^2}{1-d^2}e^{\alpha+\beta}+e^{\alpha+\beta}+e^{\frac{3\beta-\alpha}{2}},\qquad J=e^{\alpha+\beta}.
\ea
We note that  the module defined above  is analogue of the Wakimoto module for $\widehat{\mathfrak{sl}}_2$.

Using the free boson representation, we can easily compute the coefficient of $t_1^n$ as follows:
\bq
\begin{split}
&\quad\braket{\alpha,\beta |(G_0^+ )^n(G_0^-)^n|\alpha,\beta}\\
=\ \ &\prod_{i=0}^{n-1}\frac{q^{-i}e^{\beta}-q^ie^{-\beta}}{q-q^{-1}}\prod_{j=1}^n\frac{q^{-3i}e^{\alpha}-q^{-i}e^{\alpha}-q^{-i}e^{\frac{\beta-\alpha}{2}}+q^ie^{\frac{\beta-\alpha}{2}}}{(q-q^{-1})^2}\\
=\ \ &\prod_{i=1}^{n}\frac{(q^{-i+1}e^{\beta}-q^{i-1}e^{-\beta})(e^{\frac{\beta-\alpha}{2}}-q^{-2i}e^\alpha)(q^i-q^{-i})}{(q-q^{-1})^3}
\end{split}
\label{eq:Whittakernorm1}
\eq

Next, we consider the coefficient of $t_2^m$ in (\ref{eq:Wnorm}). One can see that the way using free boson representation requires a large amount of computation. In the following, we mainly use the algebraic relation. 
From (\ref{eq:GGrelation}), we have
\small
\bq
\label{eq:recursionnorm}
\begin{split}
&\braket{\alpha,\beta |(G_1^- )^n(G_{-1}^+)^n|\alpha,\beta}\\
=\ &\sum_{m=0}^{n-1}\braket{\alpha,\beta |(G_1^- )^{n-1}(G_{-1}^+)^m[G_1^-,G_{-1}^+](G_{-1}^+)^{n-m-1}|\alpha,\beta}\\
=\ &\sum_{m=0}^{n-1}\braket{\alpha,\beta |(G_1^-)^{n-1}(G_{-1}^+)^m\left(\frac{q(1-q_3^2)}{(1-d^{-2})(q-q^{-1})^2}\psi_0-\frac{1}{d^2(q-q^{-1})^2}T_0+\frac{1}{q^2d^4(q-q^{-1})^2}\tilde{\psi}^-_0\right)(G_{-1}^+)^{n-m-1}|\alpha,\beta}.
\end{split}
\eq
\normalsize
We see from this equation that we can implement the computation if we  obtain the eigenvalues of $\psi_0$, $T_0$ and $\tilde{\psi}_0$. \footnote{We note that  arbitrary descendant states are not necessarily eigenvectors of the zero modes as opposed to CFT case because the degeneracy is resolved after q-deformation. For the state  $(G_{-1}^+)^n\ket{\alpha,\beta}$, we do not need to care about that because there is no other state with the same charge. }

We begin with $\psi_0$. From (\ref{eq:psiGp}), we have
\bq
\label{eq:psi0Gp}
\begin{split}
&\psi_0G^+_{-1}-q^{-2}G^+_{-1}\psi_0+(q^2-q^{-2})\sum_{n\geq1}^{\infty}q^{-2n}G^+_{-n-1}\psi_n\\
=\ &\sum_{n\in\mathbb{Z}}\left(-\frac{(1-q^4)(1-q^{-4}d^{-2})q_2^n}{1-q_3^2}+\frac{(1-q^{-2})(1-q_1^{-2})d^{-2n}}{1-q_3^2}\right)\sum_{m\geq{\rm max}(0,-n)}^{\infty}\psi^-_{-m-n}G^+_{n-1}\psi^+_{m}.
\end{split}
\eq
When we consider the action of (\ref{eq:psi0Gp}) on $(G_{-1}^+)^l\ket{\alpha,\beta}$,  most of the terms vanish and are simplified as follows:
\bq
\begin{split}
&(\psi_0G^+_{-1}-q^{-2}G^+_{-1}\psi_0)(G_{-1}^+)^l\ket{\alpha,\beta}\\
=\ &\left(-\frac{(1-q^4)(1-q^{-4}d^{-2})}{1-q_3^2}+\frac{(1-q^{-2})(1-q_1^{-2})}{1-q_3^2}\right)\psi^-_{0}G^+_{-1}\psi^+_{0}(G_{-1}^+)^l\ket{\alpha,\beta}.
\end{split}
\eq
We can simplify the RHS as
\ba
\label{eq:psi0Gprec}
(\psi_0G^+_{-1}-q^{-2}G^+_{-1}\psi_0)(G_{-1}^+)^l\ket{\alpha,\beta}=q^{4l}(q^3-q^{-3})e^{\alpha+\beta}(G_{-1}^+)^{l+1}\ket{\alpha,\beta}.
\ea
Then we can obtain the eigenvalue of $\psi_0$ recursively from (\ref{eq:psi0Gprec}),
\bq
\label{eq:psi0eigenvalue}
\begin{split}
\psi_0(G_{-1}^+)^l\ket{\alpha,\beta}=&\left(q^{-2l-1}e^{\alpha+\beta}+(q^3-q^{-3})\sum_{i=0}^{l-1}q^{-2i}q^{4(l-i-1)}e^{\alpha+\beta}\right)(G_{-1}^+)^l\ket{\alpha,\beta}\\
=&\ q^{4l-1}e^{\alpha+\beta}(G_{-1}^+)^l\ket{\alpha,\beta}.
\end{split}
\eq
We can apply the similar discussion to $T_0$, which lead to
\bq
(q^{-2}T_0G^+_{-1}-G^+_{-1}T_0)(G_{-1}^+)^l\ket{\alpha,\beta}
=-\frac{(1-q_1^{-2})(1-q^{-2})d^2}{1-d^{-2}}q^{4l}e^{\alpha+\beta}(G_{-1}^+)^l\ket{\alpha,\beta}.
\eq
The we have
\small
\bq
\label{eq:T0eigenvalue}
\begin{split}
&\quad T_0(G_{-1}^+)^l\ket{\alpha,\beta}\\
=\ &\left(q^{2l}\bigl(q^{-2}e^{\alpha-\beta}-\frac{1-q_1^2}{1-d^2}e^{\alpha+\beta}+e^{\alpha+\beta}+e^{\frac{3\beta-\alpha}{2}}\bigr)
-\frac{(1-q_1^{-2})(1-q^{-2})q_3^{-2}}{1-d^{-2}}\sum_{i=0}^{l-1}q^{4l-2i-4}e^{\alpha+\beta}\right)(G_{-1}^+)^l\ket{\alpha,\beta}\\
=\ &\left(q^{2l-2}e^{\alpha-\beta}+\biggl(q^{2l-2}d^2-\frac{(1-q_1^2)d^2}{1-d^2}q^{4l}\biggr)e^{\alpha+\beta}+q^{2l}e^{\frac{3\beta-\alpha}{2}}\right)(G_{-1}^+)^l\ket{\alpha,\beta}.
\end{split}
\eq
\normalsize
For $\tilde{\psi}^-_0$, we have
\ba
\tilde{\psi}^-_0G^+_{-1}=q^{-2}G^+_{-1}\tilde{\psi}^-_0,
\ea
which leads to
\ba
\label{eq:psit0eigenvalue}
\tilde{\psi}^-_0(G_{-1}^+)^l\ket{\alpha,\beta}=q^{-2l}e^{-\frac{\alpha+\beta}{2}}(G_{-1}^+)^l\ket{\alpha,\beta}.
\ea
By substituting (\ref{eq:psi0eigenvalue}), (\ref{eq:T0eigenvalue}) and (\ref{eq:psit0eigenvalue}) into (\ref{eq:recursionnorm}), we have the following recursion relation, 
\bq
\begin{split}
\braket{\alpha,\beta |(G_1^- )^n(G_{-1}^+)^n|\alpha,\beta}=f_n\braket{\alpha,\beta |(G_1^- )^{n-1}(G_{-1}^+)^{n-1}|\alpha,\beta} ,\\
\end{split}
\eq
where
\bq
\begin{split}
f_n&=\frac{1}{(q-q^{-1})^2}\sum_{l=0}^{n-1}\left((q^{4l}+q^{4l-2}-q^{2l-2})e^{\alpha+\beta}-q^{2l-2}d^{-2}e^{\alpha-\beta}+q^{-2l-2}d^{-4}e^{-\frac{\alpha+\beta}{2}}-q^{2l}d^{-2}e^{\frac{3\beta-\alpha}{2}}\right)\\
&=\frac{-q^{-3}}{(q-q^{-1})^3}\left((q^{2n}-q^{4n})e^{\alpha+\beta}-d^{-2}(1-q^{2n})e^{\alpha-\beta}-q^2d^{-4}(1-q^{-2n})e^{-\frac{\alpha+\beta}{2}}-q^2d^{-2}(1-q^{2n})e^{\frac{3\beta-\alpha}{2}}\right)\\
&=\frac{-q^{-3}(1-q^{2n})(q^{2n}e^{\alpha}-q_1^{-2}e^{\frac{\beta-\alpha}{2}})(e^{\beta}-q^{-2n}d^{-2}e^{-\beta})}{(q-q^{-1})^3}.
\end{split}
\eq
Then we have
\bq
\braket{\alpha,\beta |(G_1^- )^n(G_{-1}^+)^n|\alpha,\beta}
=(-1)^n\prod_{i=1}^n\frac{q^{-3}(1-q^{2i})(q^{2i}e^{\alpha}-q_1^{-2}e^{\frac{\beta-\alpha}{2}})(e^{\beta}-q^{-2i}d^{-2}e^{-\beta})}{(q-q^{-1})^3}.
\label{eq:Whittakernorm2}
\eq

Let us compare  (\ref{eq:Whittakernorm1}) and (\ref{eq:Whittakernorm2}) with the partition function (\ref{eq:Zsurface1}) and (\ref{eq:Zsurface2}). For convenience, we denote the Omega background parameters by ($q_1',q_2'$) here.
If we set
\ba
&&q_1'=q^2,\qquad q_2'=q_1^2,\label{eq:comparison_omega}\\
&&\frac{u_{1,1}}{u_{1,2}}=e^{-2\beta}d^{-2},\qquad\frac{u_{2,1}}{u_{1,2}}=e^{\frac{\beta-3\alpha}{2}}q_1^{-2}
\ea
and tune the fugacity parameters appropriately, the instanton partition function matches with the norm of the Whittaker states up to the difference between "$1-e^X$" and "$2\sinh \frac{X}{2}$". We expect it can be resolved by introducing the Chern-Simons term\footnote{We note that 5d supersymmetric gauge theories with  surface operators and Chern-Simons terms were discussed in \cite{Ashok:2017lko,Ashok:2017bld}. }.
We note that the parameters in (\ref{eq:comparison_omega}) are exactly the same as those of the subalgebra $\mathcal{E}_1^{(1)}$ given in (\ref{eq:upper}).

\section{Conclusion and Discussion}
\label{sec:conclusion}
In this paper, we proposed the quantum deformation of Feigin-Semikhatov's W-algebras. We first studied the simplest case $U_q(\widehat{\mathfrak{sl}}_2)$ in terms of the gluing construction proposed in \cite{Gaiotto:2017euk,Prochazka:2017qum}. We found that the currents $\mathscr{E}_2(z),\mathscr{F}_2(z)$ can be expressed as the product of the vertex operators for q-$W(\widehat{\mathfrak{gl}}_{2|1})$ and q-$W(\widehat{\mathfrak{gl}}_1)$. Based on this result, we constructed the vertex operators for q-$W(\widehat{\mathfrak{gl}}_{N|1})$ by demanding the commutativity with the screening charges proposed in \cite{bershtein2018plane}. We fixed the ambiguity for the Heisenberg part by requiring the Weyl transformation which flips the Dynkin diagram of $\mathfrak{gl}_{n|1}$. We also checked the relation to 5d AGT correspondence with a simple surface operator for the deformed Bershadsky-Polyakov algebra.

There are several directions for the future work.
\begin{itemize}
\item It is interesting to extend the discussion of this paper to another (super) W-algebra. In the case of Feigin-Semikhatov's W-algebras, one of $\mathcal{E}_1$ used in the gluing construction is  merely Heisenberg algebra, which allows us to assume the gluing currents $\mathscr{E}_N(z),\mathscr{F}_N(z)$ to be the product of the two vertex operators. It would be not true for the another W-algebra and we may need alternative method. 
\item In this paper, we use free field representation in the gluing construction, but it may be possible to do the same process at the level of the quantum toroidal algebra. In fact, it is expected that Gaiotto-Rap\v{c}\'{a}k's VOA at least for a certain brane-web is a truncation of a quantum toroidal algebra \cite{Rapcak:2019abg}. We expect that the deformed Feigin-Semikhatov's W-algebras can be realized from the shifted toroidal algebra of $gl_2$ defined in \cite{Negut:2019agq}.
\item There is a close relation between 2d CFT and integrable systems \cite{Bazhanov:1994ft,Bazhanov:1996dr,Bazhanov:1998dq}.  It was also studied in the q-deformed case  \cite{FKSW1,FKSW2} and has been recently studied in terms of the quantum toroidal algebra \cite{Feigin:2015raa,Feigin:2017wnq,Feigin:2017caw,Feigin:2017gcv,Feigin:2020glm}. It would be interesting to study the integrals of motion for the deformed Feigin-Semikhatov's W-algebras using the quantum toroidal algebras. It would be also interesting to explore the relation to symmetric functions.
\item The insertion of a simple surface operator can be realized in two ways in terms of 6d theory. If we consider a codimension 4 operator instead of condimension 2, it does not change the dual algebra but serves as the insertion of degenerate primary fields of $W_N$ \cite{Alday:2009fs,Kozcaz:2010af}. The relation between the correlation functions for different W-algebras has been recently studied  well in   \cite{Creutzig:2020ffn}.  We hope that the relation is helpful to define q-analogue of the degenerate primary fields of $W_N$. 

\end{itemize}

\section*{Acknowledgement}
The author would like to thank H. Awata, M. Fukuda, H. Kanno, Y. Matsuo and J. Shiraishi for helpful discussions and comments. He is partially supported by JSPS fellowship.

\appendix
\section{The free boson representation of $\mathcal{E}_1$ from the deformed Wakimoto representation}
\label{app:e1e0}
In this section, we use the following relations for the Wakimoto bosons:
\bq
\begin{split}
{\rm for\ } n>0\qquad&[u_{1,n}^+,\hat{u}_{1,m}^+]=[u_{1,n}^-,\hat{u}_{1,m}^-]=nd^n(q^n-q^{-n})\delta_{n+m,0},\qquad [u_{1,n}^-,\hat{u}_{1,m}^+]=0,\\
&[u_{1,n}^+,\hat{u}_{1,m}^-]=n(q^n-q^{-n})(d^n+d^{-n})\delta_{n+m,0},\\
\end{split}
\eq
\bq
[a_{2,0},Q_{u_1}]=1,\qquad [a_{3,0},Q_{u_1}]=-1.\hspace{50pt}
\eq
The free boson representation of the current (\ref{eq:10current}) is given as follows:
\bq
\begin{split}
&\rho(E_{1|0}^{(1)}(z))\\
=\ &\lim_{z'\to z}\frac{(1-\frac{z}{z'})(1-\frac{q_3z}{q_1z'})}{(q-q^{-1})^2}\left(\frac{1-\frac{q^2z}{z'}}{q(1-\frac{z}{z'})}:e^{u_1^+(q_1z')+\hat{u}_1^+(z)}:+\frac{(1-\frac{z}{q_1^2z'})(1-\frac{d^2z}{z'})}{q(1-\frac{z}{z'})(1-\frac{z}{d^2z'})}:e^{u_1^+(q_1z')+\hat{u}_1^-(z)}:\right.\\
&\hspace{280pt}\left.+\frac{q(1-\frac{q^2z}{z'})}{1-\frac{z}{z'}}:e^{u_1^-(q_1z')+\hat{u}_1^-(z)}:\right), \\
=\ &-\frac{1-d^{-2}}{q-q^{-1}}:e^{u_1^+(q_1z')+\hat{u}_1^+(z)}:+\frac{(1-q_1^{-2})(1-d^2)}{q(q-q^{-1})^2}:e^{u_1^+(q_1z)+\hat{u}_1^-(z)}:-\frac{q^2(1-d^{-2})}{q-q^{-1}}:e^{u_1^-(q_1z')+\hat{u}_1^-(z)}:
\end{split}
\eq
\bq
\begin{split}
\rho(E_{1|0}^{(3)}(z))&=\lim_{z'\to z}\frac{(1-\frac{z}{z'})(1-\frac{q_1z}{q_3z'})}{(q-q^{-1})^2}\frac{(1-\frac{d^4z}{z'})(1-\frac{q^2z}{z'})}{q(1-\frac{z}{z'})(1-\frac{d^2z}{z'})}:e^{u_1^+(q_3z')+\hat{u}_1^+(z)}:\hspace{30pt}\\
&=-\frac{1-d^4}{q-q^{-1}}:e^{u_1^+(q_3z)+\hat{u}_1^+(z)}:.
\end{split}
\eq
The coefficients do not matter because we can absorb them into the zero mode.


\section{Useful formulae}
\label{app:formulae}

The commutation relations among the oscillators $A_n,B_n,C_n,D_n$ are given as follows:
\bq
\label{eq:ABCDcom}
\begin{split}
{\rm for\  }n\geq0,\ \ &[A_n,A_m]=[B_n,B_m]=[C_n,C_m]=[D_n,D_m]=-n(q^n-q^{-n})q^n\delta_{n+m,0},\\
&[A_n,B_m]=[A_n,C_m]=[B_n,D_m]=[C_n,D_m]=-n(q^{2n}-q^{-2n})\delta_{n+m,0},\\
&[C_n,B_m]=-n(q^n-q^{-n})q_1^nd^n\delta_{n+m,0},\\
&[D_n,A_m]=n(q^n-q^{-n})q^nd^{2n}\delta_{n+m,0},\\
&[A_n,D_m]=-n(q^n-q^{-n}))(q^n+q^{-n}+q^{-n}d^{-2n})\delta_{n+m,0},\\
&[B_n,C_m]=-n(q^n-q^{-n}))(q^n+q^{-n}-q^nd^{-2n})\delta_{n+m,0},\\
&[B_n,A_m]=[C_n,A_m]=[D_n,B_m]=[D_n,C_m]=0,
\end{split}
\eq
\bq
\begin{split}
{\rm for\  }n\in\mathbb{Z},\ \ &[A_n,v_{1,m}^+]=n(q^n-q^{-n})q^n(d^{-n}+d^{-3n})\delta_{n+m,0},\hspace{120pt}\\
&[B_n,v_{1,m}^+]=[C_n,v_{1,m}^+]=n(q^n-q^{-n})q^nd^{-n}\delta_{n+m,0},\\
&[A_n,v_{1,m}^-]=n(q^n-q^{-n})q^nd^{-n}\delta_{n+m,0},\\
&[D_n,v_{1,m}^-]=-n(q^n-q^{-n})q^nd^n\delta_{n+m,0},\\
&[D_n,v_{1,m}^+]=[B_n,v_{1,m}^-]=[C_n,v_{1,m}^-]=0.
\end{split}
\eq
We can extend all of the relations (\ref{eq:ABCDcom}) to $n\in\mathbb{Z}$ as follows:
\ba
\label{eq:XYABCD}
[Y_n,Z_m]=({\rm RHS\ of\ (\ref{eq:ABCDcom})})+n\theta(n<0)(q^{2n}-q^{-2n})\delta_{n+m,0}\quad(Y,Z\in{\{A,B,C,D\}}),
\ea
where $\theta(n<0)=\begin{cases}1\ (n <0)\\0\ (n\geq0)\end{cases}$. Due to the additional term in (\ref{eq:XYABCD}), we need a rational factor in the quadratic relations in order to produce the delta function $\delta(\frac{w}{z})$.
The contractions between the two vertex operators are given as follows:
\small
\bq
\begin{split}
&\mathscr{K}(z)e^{v_1^+(w)}=q^{-4}\frac{(1-\frac{w}{q_3^2z})(1-\frac{q^2w}{z})}{(1-\frac{d^2w}{z})(1-\frac{w}{q^2z})}:\mathscr{K}(z)e^{v_1^+(w)}:,\quad \mathscr{K}(z)e^{v_1^-(w)}=q^{-4}\frac{(1-\frac{w}{q_3^2z})(1-\frac{q^2w}{z})}{(1-\frac{d^2w}{z})(1-\frac{w}{q^2z})}:\mathscr{K}(z)e^{v_1^-(w)}:,\\
&e^{v_1^+(w)}\mathscr{K}(z)=\frac{1-\frac{q_3^2z}{w}}{1-\frac{z}{d^2w}}:e^{v_1^+(w)}\mathscr{K}(z):,\quad e^{v_1^-(w)}\mathscr{K}(z)=\frac{1-\frac{q_3^2z}{w}}{1-\frac{z}{d^2w}}:e^{v_1^-(w)}\mathscr{K}(z):,\\
&\tilde{\mathscr K}^-(z)e^{v_1^+(w)}=q^2:\tilde{\mathscr K}^-(z)e^{v_1^+(w)}:,\quad\tilde{\mathscr K}^-(z)e^{v_1^-(w)}=q^2:\tilde{\mathscr K}^-(z)e^{v_1^-(w)}:,\\
&e^{v_1^+(w)}\tilde{\mathscr K}^-(z)=\frac{1-\frac{q^2z}{w}}{1-\frac{z}{q^2w}}:e^{v_1^+(w)}\tilde{\mathscr K}^-(z):,\quad e^{v_1^-(w)}\tilde{\mathscr K}^-(z)=\frac{1-\frac{q^2z}{w}}{1-\frac{z}{q^2w}}:e^{v_1^-(w)}\tilde{\mathscr K}^-(z):,\\
&\mathscr{K}(z)e^{X(w)}=q^4\frac{(1-\frac{q^{-2}dw}{z})(1-\frac{q^{-2}d^3w}{z})}{(1-\frac{dw}{z})(1-\frac{q^2d^3w}{z})}:\mathscr{K}(z)e^{X(w)}:,\quad(X=A,B,C,D)\\
&e^{X(w)}\mathscr{K}(z)=\frac{1-\frac{z}{q^{-2}dw}}{1-\frac{z}{dw}}:e^{X(w)}\mathscr{K}(z):,\quad(X=A,B,C,D)\\
&\tilde{\mathscr K}^-(z)e^{X(w)}=q^{-2}:\tilde{\mathscr K}^-(z)e^{X(w)}:\quad(X=A,B,C,D)\\
&e^{X(w)}\tilde{\mathscr K}^-(z)=\frac{1-\frac{dz}{w}}{1-\frac{q^4dz}{w}}:e^{X(w)}\tilde{\mathscr K}^-(z):,\quad (X=A,B,C,D)\\
&\mathscr{X}(z)e^{v^+_1(w)}=-\frac{q^{-4}(1-\frac{w}{q_1^2z})(1-\frac{q^2w}{z})}{(1-\frac{w}{d^2z})(1-\frac{w}{q^2z})}\frac{	1-q_1^2}{1-d^2}:e^{A(d^{-1}z)+v^+_1(z)+v^+(w)}:+\frac{q^{-2}(1-\frac{q^2w}{z})}{1-\frac{w}{q^2z}}:e^{B(d^{-1}z)+v^+(z)+v^-(w)}:\\
&\hspace{65pt}+q^{-2}\frac{1-\frac{q^2w}{z}}{1-\frac{w}{q^2z}}:e^{C(d^{-1}z)+v^+_1(z)+v^+_1(w)}:+q^{-4}\frac{(1-\frac{q^2w}{z})(1-\frac{w}{q_1^2z})}{(1-\frac{w}{z})(1-\frac{w}{d^2z})}:e^{A(d^{-1}z)+v^-_1(z)+v^+(w)}:\\
&\mathscr{X}(z)e^{v^-_1(w)}=q^{-4}\frac{(1-\frac{q^2w}{z})^2}{(1-\frac{w}{z})(1-\frac{w}{q^2z})}:e^{A(d^{-1}z)+v^+_1(z)+v^-(w)}:\\
&\hspace{65pt}-q^{-2}\frac{1-\frac{q^2w}{z}}{1-\frac{w}{q^2z}}(:e^{B(d^{-1}z)+v_1^+(z)+v_1^-(w)}:+:e^{C(d^{-1}z)+v_1^+(z)+v_1^-(w)}:+q^{-2}:e^{A(d^{-1}z)+v_1^+(z)+v_1^-(w)}:)
\end{split}
\eq
\normalsize

For generic $\mathcal{W}_N^{(2)}$, the contractions are given as follows:
\bq
\begin{split}
&\mathscr{E}_N^{s_3,s_4\cdots s_{N+1}}(z)e^{v_1^+(w)}\\
=\ &q^{2-N-\sum_{i=3}^{N+1}s_i}\exp\left(\sum_{n>0}\frac{(1-q^{-2n})(d^{(N-3)n}-d^{-(2+\sum_{i=3}^{N+1}s_i )n})}{n(d^n-d^{-n})}\biggl(\frac{w}{z}\biggr)^n\right):\mathscr{E}_N^{s_3,s_4\cdots s_{N+1}}(z)e^{v_1^+(w)}:,
\end{split}
\label{eq:Ncontraction1}
\eq
\bq
\begin{split}
&e^{v_1^+(w)}\mathscr{E}_N^{s_3,s_4\cdots s_{N+1}}(z)\\
=\ &q\exp\left(-\sum_{n>0}\frac{(1-q^{2n})(d^{(3-N)n}-d^{(2+\sum_{i=3}^{N+1}s_i) n})}{n(d^n-d^{-n})}\biggl(\frac{z}{w}\biggr)^n\right):\mathscr{E}_N^{s_3,s_4\cdots s_{N+1}}(z)e^{v_1^+(w)}:,\hspace{60pt}
\end{split}
\label{eq:Ncontraction2}
\eq

\bq
\begin{split}
&\mathscr{E}_N^{s_3,s_4\cdots s_{N+1}}(z)e^{v_1^-(w)}\\
=\ &q^{2-N-\sum_{i=3}^{N+1}s_i}\exp\left(\sum_{n>0}\frac{(1-q^{-2n})(d^{(N-3)n}-d^{-\sum_{i=3}^{N+1}s_i n})}{n(d^n-d^{-n})}\biggl(\frac{w}{z}\biggr)^n\right):\mathscr{E}_N^{s_3,s_4\cdots s_{N+1}}(z)e^{v_1^-(w)}:,
\end{split}
\label{eq:Ncontraction3}
\eq

\bq
\begin{split}
&e^{v_1^-(w)}\mathscr{E}_N^{s_3,s_4\cdots s_{N+1}}(z)\\
=\ &q^{-1}\exp\left(-\sum_{n>0}\frac{(1-q^{2n})(d^{(3-N)n}-d^{\sum_{i=3}^{N+1}s_i n})}{n(d^n-d^{-n})}\biggl(\frac{z}{w}\biggr)^n\right):\mathscr{E}_N^{s_3,s_4\cdots s_{N+1}}(z)e^{v_1^-(w)}:,\hspace{60pt}
\end{split}
\label{eq:Ncontraction4}
\eq
From (\ref{eq:Ncontraction1}) and (\ref{eq:Ncontraction2}), we have
\bq
\begin{split}
&[\mathscr{E}_N^{s_3,s_4\cdots s_{N+1}}(z),e^{v_1^+(w)}]\\
=\ &\begin{cases}
\quad0\qquad\qquad (s_3=s_4=\cdots=s_{N+1}=-1)\\
q^{2-N-\sum_{i=3}^{N+1}s_i}(1-q^2)\displaystyle{\sum_{i=0}^{\frac{N-3+\sum_{k=3}^{N+1}s_k}{2}}}\delta\left(\frac{d^{N-4-2i}}{z}\right)\Biggl(\prod_{\substack{j=0\\ j\neq i}}^{\frac{N-3+\sum_{k=3}^{N+1}s_k}{2}}\frac{1-q^2d^{2(i-j)}}{1-d^{2(i-j)}}\Biggr):\mathscr{E}_N^{s_3,s_4\cdots s_{N+1}}(z)e^{v_1^+(w)}:.
\end{cases}\\
&\hspace{400pt}({\rm otherwise})
\end{split}
\label{eq:Ncommutator1}
\eq
From (\ref{eq:Ncontraction3}) and (\ref{eq:Ncontraction4}), we have
\bq
\begin{split}
&[\mathscr{E}_N^{s_3,s_4\cdots s_{N+1}}(z),e^{v_1^-(w)}]\\
=\ &\begin{cases}
(q-q^{-1})\delta\left(\frac{q^2d^{N-2}w}{z}\right):\mathscr{E}_N^{s_3,s_4,\cdots,s_{N+1}}(z)e^{v_1^-(w)}:\qquad(s_3=s_4=\cdots=s_{N+1}=-1)\\
\quad 0\qquad (\#\{i\in\{3,4\cdots N+1\}|s_i=1\}=1)\\
q^{2-N-\sum_{i=3}^{N+1}s_i}(1-q^2)\displaystyle{\sum_{i=0}^{\frac{N-5+\sum_{k=3}^{N+1}s_k}{2}}\delta\left(\frac{d^{N-4-2i}}{z}\right)\Biggl(\prod_{\substack{j=0\\ j\neq i}}^{\frac{N-5+\sum_{k=3}^{N+1}s_k}{2}}\frac{1-q^2d^{2(i-j)}}{1-d^{2(i-j)}}\Biggr)}:\mathscr{E}_N^{s_3,s_4\cdots s_{N+1}}(z)e^{v_1^-(w)}:.
\end{cases}\\
&\hspace{400pt}({\rm otherwise})
\end{split}
\label{eq:Ncommutator2}
\eq

\section{The proof of (\ref{eq:ENEN})}
\label{app:proof}
We introduce the following notation:
\ba
\label{eq:LambdaLambda}
\Lambda(z)\Lambda(w)=f\left(\frac{w}{z}\right):\Lambda(z)\Lambda(w):,\quad f\left(\frac{w}{z}\right)=\exp\left(\sum_{n>0}\frac{(q^n-q^{-n})q_3^{-n}}{n(d^n-d^{-n})}\biggl(\frac{w}{z}\biggr)^n\right).
\ea
Then (\ref{eq:ENEN}) is equivalent to 
\ba
\label{eq:tENtEN}
(z-q_2w)f\left(\frac{w}{z}\right)\mathscr{E'}_N(z)\mathscr{E'}_N(w)-(w-q_2z)f\left(\frac{z}{w}\right)\mathscr{E'}_N(w)\mathscr{E'}_N(z)=0.
\ea
Let us show (\ref{eq:tENtEN}) by induction.
 We assume (\ref{eq:tENtEN}) is true for $N$. When we express the quadratic relation for $\mathscr{E'}_{N+1}(z)$ by $\mathscr{E'}_N(z)$ using (\ref{eq:recursionE}), it is expanded into the four parts. The first one is 
\bq
\begin{split}
&(z-q_2w)\mathscr{E'}_N(dz)e^{\phi_{(N+1)}^+(z)}\Lambda(d^{2-N}z)\mathscr{E'}_N(dw)e^{\phi^+_{(N+1)}(w)}\Lambda(d^{2-N}w)-(z\leftrightarrow w)\\
=\ &q_2(z-q_2w)f\left(\frac{w}{z}\right)\mathscr{E'}_N(dz)\mathscr{E'}_N(dw)e^{\phi_{(N+1)}^+(z)}e^{\phi^+_{(N+1)}(w)}:\Lambda(d^{2-N}z)\Lambda(d^{2-N}w):-(z\leftrightarrow w)\\
=\ &0,
\end{split}
\eq
The second one is 
\bq
\begin{split}
&(z-q_2w)\mathscr{E'}_N(dz)e^{\phi_{(N+1)}^+(z)}\Lambda(d^{2-N}z)e^{\phi_{(N+1)}^-(w)}\mathscr{E'}_N(d^{-1}w)\Lambda(d^{2-N}w)\\
&\hspace{80pt}-(w-q_2z)e^{\phi_{(N+1)}^-(w)}\mathscr{E'}_N(d^{-1}w)\Lambda(d^{2-N}w)\mathscr{E'}_N(dz)e^{\phi_{(N+1)}^+(z)}\Lambda(d^{2-N}z)\\
=\ &q_2(z-q_2w)\mathscr{E'}_N(dz)\mathscr{E'}_N(d^{-1}w)\frac{(1-\frac{w}{q_2z})(1-\frac{w}{q_1^2z})}{(1-\frac{w}{z})(1-\frac{w}{d^2z})}e^{\phi_{(N+1)}^-(w)}e^{\phi_{(N+1)}^+(z)}f\left(\frac{w}{z}\right):\Lambda(d^{2-N}z)\Lambda(d^{2-N}w):\\
&\hspace{80pt}-(w-q_2z)\mathscr{E'}_N(d^{-1}w)\mathscr{E'}_N(dz)e^{\phi_{(N+1)}^-(w)}e^{\phi_{(N+1)}^+(z)}f\left(\frac{z}{w}\right):\Lambda(d^{2-N}w)\Lambda(d^{2-N}z):\\
=\ & q_2\frac{1-\frac{w}{q_2z}}{1-\frac{w}{d^2z}}(z-q_1^{-2}w)f\left(\frac{w}{d^2z}\right)\mathscr{E'}_N(dz)\mathscr{E'}_N(d^{-1}w)e^{\phi_{(N+1)}^-(w)}e^{\phi_{(N+1)}^+(z)}:\Lambda(d^{2-N}z)\Lambda(d^{2-N}w):\\
&\hspace{40pt}-\frac{1-\frac{q_2z}{w}}{1-\frac{d^2z}{w}}(w-q_3^{-2}z)f\left(\frac{d^2z}{w}\right)\mathscr{E'}_N(d^{-1}w)\mathscr{E'}_N(dz)e^{\phi_{(N+1)}^-(w)}e^{\phi_{(N+1)}^+(z)}:\Lambda(d^{2-N}w)\Lambda(d^{2-N}z):\\
=\ &q_2(1-q_1^2)\delta\left(\frac{w}{d^2z}\right)\Biggl((z-q_1^{-2}w)f\left(\frac{w}{d^2z}\right)\mathscr{E'}_N(dz)\mathscr{E'}_N(d^{-1}w)\Biggr)e^{\phi_{(N+1)}^-(w)}e^{\phi_{(N+1)}^+(z)}:\Lambda(d^{2-N}z)\Lambda(d^{2-N}w):\\
=\ &0.
\end{split}
\eq
To obtain the last line, we use the relation
\ba
(z-q_2w)f\left(\frac{w}{z}\right)\mathscr{E'}_N(z)\mathscr{E'}_N(w)\propto z-w
\ea
which is derived from (\ref{eq:tENtEN}). The third one is 
\bq
\begin{split}
&(z-q_2w)e^{\phi_{(N+1)}^-(z)}\mathscr{E'}_N(d^{-1}z)\Lambda(d^{2-N}z)\mathscr{E'}_N(dw)e^{\phi_{(N+1)}^+(w)}\Lambda(d^{2-N}w)\\
&\hspace{80pt}-(w-q_2z)\mathscr{E'}_N(dw)e^{\phi_{(N+1)}^+(w)}\Lambda(d^{2-N}w)e^{\phi_{(N+1)}^-(z)}\mathscr{E'}_N(d^{-1}z)\Lambda(d^{2-N}z)\\
=\ &\frac{1-\frac{q_2w}{z}}{1-\frac{d^2w}{z}}(z-q_3^{-2}w)\mathscr{E'}_N(d^{-1}z)\mathscr{E'}_N(dw)e^{\phi_{(N+1)}^-(z)}e^{\phi_{(N+1)}^+(w)}:\Lambda(d^{2-N}z)\Lambda(d^{2-N}w):\\
&\hspace{80pt}-q_2\frac{1-\frac{z}{q_2w}}{1-\frac{z}{d^2w}}(w-q_1^{-2}z)\mathscr{E'}_N(dw)\mathscr{E'}_N(d^{-1}z)e^{\phi_{(N+1)}^+(w)}e^{\phi_{(N+1)}^-(z)}:\Lambda(d^{2-N}w)\Lambda(d^{2-N}z):\\
=\ &(1-q_1^{-2})\delta\left(\frac{d^2w}{z}\right)\Biggl((z-q_3^{-2}w)f\left(\frac{d^2w}{z}\right)\mathscr{E'}_N(d^{-1}z)\mathscr{E'}_N(dw)\Biggr)e^{\phi_{(N+1)}^-(z)}e^{\phi_{(N+1)}^+(w)}:\Lambda(d^{2-N}z)\Lambda(d^{2-N}w):\\
=\ &0.
\end{split}
\eq
The last one is 
\bq
\begin{split}
&(z-q_2w)e^{\phi_{(N+1)}^-(z)}\mathscr{E'}_N(d^{-1}z)\Lambda(d^{2-N}z)e^{\phi_{(N+1)}^-(w)}\mathscr{E'}_N(d^{-1}w)\Lambda(d^{2-N}w)-(z\leftrightarrow w)\\
=\ &(z-q_2w)f\left(\frac{w}{z}\right)\mathscr{E'}_N(d^{-1}z)\mathscr{E'}_N(d^{-1}w)e^{\phi_{(N+1)}^-(z)}e^{\phi^-_{(N+1)}(w)}:\Lambda(d^{2-N}z)\Lambda(d^{2-N}w):-(z\leftrightarrow w)\\
=\ &0.
\end{split}
\eq
From the above, the relation (\ref{eq:ENEN}) holds also for $N+1$.\hspace{150pt} $\Box$

\bibliography{qFS}
 
\end{document}